\documentclass[]{pasj02} 
\usepackage[switch,mathlines]{lineno} 

\usepackage{xcolor}
\usepackage{graphicx}
\usepackage{threeparttable}

\jyear{2025}
\Received{2025/05/29}
\Accepted{2025/07/30}

\newcommand{\chandra}{\textrm{Chandra}}
\newcommand{\xmmn}{\textrm{XMM-Newton}}
\newcommand{\xmm}{\textrm{XMM}}

\newcommand{\xrism}{\textrm{XRISM}}

\begin{document} 

\title{XRISM Observations of Cassiopeia A: Overview, Atomic Data, and Spectral Models}

\author{
Paul \textsc{Plucinsky},\altaffilmark{1}\orcid{0000-0003-1415-5823}
\email{pplucinsky@cfa.harvard.edu}
Manan \textsc{Agarwal},\altaffilmark{2}\orcid{0000-0001-6965-8642}
\email{m.agarwal@uva.nl}
Liyi \textsc{Gu},\altaffilmark{3}\orcid{0000-0001-9911-7038}
Adam \textsc{Foster},\altaffilmark{1}\orcid{0000-0003-3462-8886}
Toshiki \textsc{Sato},\altaffilmark{4}\orcid{0000-0001-9267-1693}
Aya \textsc{Bamba},\altaffilmark{5,6,7}\orcid{0000-0003-0890-4920}
Jacco \textsc{Vink},\altaffilmark{2,3}\orcid{0000-0002-4708-4219}
Masahiro \textsc{Ichihashi},\altaffilmark{5}\orcid{0000-0001-7713-5016}
Kai \textsc{Matsunaga},\altaffilmark{8}\orcid{0009-0003-0653-2913}
Koji \textsc{Mori},\altaffilmark{9}\orcid{0000-0002-0018-0369}
Hiroshi \textsc{Nakajima},\altaffilmark{10}\orcid{0000-0001-6988-3938}
Frederick \textsc{Porter},\altaffilmark{11}\orcid{0000-0002-6374-1119}
Haruto \textsc{Sonoda},\altaffilmark{5,13}\orcid{0009-0006-0015-4132} 
Shunsuke \textsc{Suzuki},\altaffilmark{12,13}\orcid{0009-0008-1853-6379} 
Dai \textsc{Tateishi},\altaffilmark{5}\orcid{0000-0003-0248-4064}
Yukikatsu \textsc{Terada},\altaffilmark{14,13}\orcid{0000-0002-2359-1857}
Hiroyuki \textsc{Uchida},\altaffilmark{8}\orcid{0000-0003-1518-2188}
Hiroya \textsc{Yamaguchi}\altaffilmark{13}\orcid{0000-0002-5092-6085}
}


\altaffiltext{1}{Center for Astrophysics, Harvard-Smithsonian, MA 02138, USA}
\altaffiltext{2}{Anton Pannekoek Institute/GRAPPA, University of Amsterdam, Science Park 904, 1098 XH Amsterdam, The Netherlands}
\altaffiltext{3}{SRON Netherlands Institute for Space Research, Niels Bohrweg 4, 2333 CA Leiden, The Netherlands}
\altaffiltext{4}{School of Science and Technology, Meiji University, Kanagawa, 214-8571, Japan}
\altaffiltext{5}{Department of Physics, Graduate School of Science,
The University of Tokyo, 7-3-1 Hongo, Bunkyo-ku, Tokyo 113-0033, Japan}
\altaffiltext{6}{Research Center for the Early Universe, School of Science, The University of Tokyo, 7-3-1
Hongo, Bunkyo-ku, Tokyo 113-0033, Japan}
\altaffiltext{7}{Trans-Scale Quantum Science Institute, The University of Tokyo, Tokyo  113-0033, Japan}
\altaffiltext{8}{Department of Physics, Kyoto University, Kyoto 606-8502, Japan}
\altaffiltext{9}{Faculty of Engineering, University of Miyazaki, Miyazaki 889-2192, Japan}
\altaffiltext{10}{College of Science and Engineering, Kanto Gakuin University, Kanagawa 236-8501, Japan}
\altaffiltext{11}{NASA Goddard Space Flight Center, Greenbelt, MD 20771, USA}
\altaffiltext{12}{Department of Science and Engineering, Graduate School of Science and Engineering, Aoyama Gakuin University, 5-10-1, Fuchinobe, Sagamihara 252-5258, Japan}
\altaffiltext{13}{ISAS/JAXA, 3-1-1 Yoshinodai, Chuo-ku, Sagamihara, Kanagawa 252-5210, Japan}
\altaffiltext{14}{Graduate School of Science and Engineering, Saitama University, 255 Shimo-Ohkubo, Sakura, Saitama 338-8570, Japan}


\KeyWords{%
ISM: supernova remnants ---
ISM: individual objects (Cassiopeia A) ---
shock waves ---
supernovae: individual (Cassiopeia A)
}  

\maketitle

\begin{abstract}
Cassiopeia~A (Cas~A) is the youngest known core-collapse supernova remnant (SNR) in the Galaxy and is perhaps the best-studied SNR in X-rays. Cas~A has a line-rich spectrum dominated by thermal emission and given its high flux, it is an appealing target for high-resolution X-ray spectroscopy.
Cas~A was observed at two different locations during the Performance Verification phase of the \xrism\, mission, one location in the southeastern part (SE) of the remnant and one in the northwestern part (NW).  This paper serves as an overview of these observations and discusses some of the issues relevant for the analysis of the data.   We present maps of the so-called ``spatial-spectral mixing'' effect due to the fact that the \xrism\, point-spread function is larger than a pixel in the Resolve calorimeter array. We analyze spectra from two bright, on-axis regions such that the effects of spatial-spectral mixing are minimized.
We fit these spectra with a semi-empirical model consisting of two thermal components, a nonthermal component, and a background model. We find that it is critical to include redshifts/blueshifts and broadening of the emission lines in the two thermal components to achieve a reasonable fit given the high spectral resolution of the Resolve calorimeter. We fit the spectra with two versions of the {\texttt{AtomDB}} atomic database (3.0.9 and 3.1.0) and two versions of the {\texttt{SPEX}} (3.08.00 and 3.08.01$^*$) spectral fitting software.  We report the significant differences in the fitted parameters so that users might understand which results are sensitive to the atomic data version. Overall we find good agreement between {\texttt{AtomDB}}~3.1.0 and {\texttt{SPEX}~3.08.01$^*$} for the spectral models considered in this paper.
The most significant difference we found between {\texttt{AtomDB}}~3.0.9 and 3.1.0 and between {\texttt{AtomDB}}~3.1.0 and {\texttt{SPEX}~3.08.01$^*$} is the Ni abundance, with the new atomic data favoring a considerably lower (up to a factor of 3) Ni abundance compared to the previous versions. 
Both regions exhibit significantly enhanced abundances compared to Solar values indicating that supernova ejecta dominate the emission in these regions.
We find that the abundance ratios of Ti/Fe, Mn/Fe, \& Ni/Fe are significantly lower in the NW than the SE, with the Ti/Fe and Mn/Fe ratios consistent with zero in the NW. These different abundance ratios from regions on opposite sides of the remnant strengthen the case for an asymmetric explosion of the progenitor. We describe the semi-empirical models that were developed and suggest that they might be useful in the calibration of moderate spectral resolution instruments.

\end{abstract}

\pagewiselinenumbers 
\nolinenumbers

\section{Introduction}

Cassiopeia~A (Cas~A) is the youngest known core-collapse supernova remnant in the Galaxy with an estimated age of $\sim350$~yr \citep{thorstensen2001,fesen2006} and a distance of $\sim3.4$~kpc \citep{reed1995}.  The progenitor has been determined to be a Type~IIb supernova (SN) from light echo spectroscopy \citep{krause2008,rest2011} with an estimated mass of $15-25~M_\odot$ \citep{chevalier2003,hwang2012,lee2014}. The ejecta mass has been estimated to be $\sim3~M_\odot$ \citep{laming2003} of which the majority has been heated by the reverse shock to X-ray emitting temperatures \citep{delaney2014,laming2020}. The ejecta structure is asymmetric as the Fe ejecta are at larger radii from the presumed explosion center than the Si ejecta in some regions \citep{hughes2000,sato2021,tsuchioka2022}  and the Ti and Fe ejecta have strikingly different spatial distributions \citep{grefenstette2014}.  The prominent jet-like structures in the northeast and southwest seen in the optical and X-rays \citep{fesen2016,hwang2004} also indicate an asymmetric explosion. Three-dimensional (3D) hydrodynamical models of a neutrino-driven SN explosion are able to reproduce the asymmetric morphology of the ejecta \citep{orlando2016,orlando2021}, both the ring-like spatial structures and the inversion of the high Z and low Z material.

Cas~A has the highest X-ray flux in the 2.0-10.0~keV bandpass of any Galactic SNR with a spectrum dominated by thermal emission.  The X-ray spectrum contains bright lines of Mg, Si, S, Ar, Ca, Fe \& Ni \citep{willingale2002} produced both by ejecta heated by the reverse shock \citep{hwang2012} and circumstellar material heated by the forward shock \citep{vink2024}. Although the thermal component dominates the overall flux, the non-thermal component is a significant contributor \citep{helder2008} and has been shown to vary on timescales of years \citep{patnaude2011,uchiyama2008}. 
The energy shifts of the bright Si-K, S-K, \& Fe-K lines have been used to construct maps of the Doppler velocities \citep{willingale2002,lazendic2006,delaney2010}, with values ranging from $-2,500~{\mathrm {km~ s^{-1}}}$ to $+4,000~{\mathrm {km~ s^{-1}}}$, to explore the three-dimensional structure of the remnant.
These maps show a structure in the Fe-K emission that is predominantly blue-shifted in the southeast (SE) and predominantly red-shifted in the northwest (NW). The Si-K and S-K emission show a similar pattern of blueshifts and redshifts between the SE and NW, but with a more complicated structure possibly indicating that more velocity components are present for Si and S. Studies of the optically emitting ejecta \citep{milisavljevic2013} reveal a torus-like geometry tilted at $30^{\circ}$ with respect to the plane of the sky with velocities ranging from $-4,000~{\mathrm {km~ s^{-1}}}$ to $+6,000~{\mathrm {km~ s^{-1}}}$.   Cas~A exhibits a complex morphology with multiple plasmas with different conditions (temperature, density, elemental abundance, ionization timescale, velocity, etc.) contributing to the X-ray emission along any given line-of-sight.

Cas~A is an appealing target for high resolution X-ray spectroscopy given its high flux and line-rich spectrum. The X-ray Imaging and Spectroscopy Mission (\xrism, \cite{tashiro2020,tashiro2024,tashiro2025}) launched on 7 September 2023 carrying two
instruments for X-ray observations.  One instrument called Resolve \citep{ishisaki2022,porter2024} is an X-ray calorimeter array with 36 pixels arranged in a $6\times6$ array covering $3\arcmin\times3\arcmin$ on the sky that provides high-resolution, non-dispersive spectroscopy (FWHM $\sim4.5$~eV at 6.0~keV) in the 1.5-12.0~keV bandpass. The other instrument called Xtend \citep{mori2022,uchida2025,noda2025} is a charge-coupled device (CCD) instrument with a large field-of-view 
($38\farcm5\times38\farcm5$) that provides moderate resolution spectroscopy in the 0.4-13.0~keV bandpass.
The \xrism\, X-ray Mirror Assembly (XMA) has an angular resolution of $1\farcm3$ half-power diameter (HPD) and $8\arcsec$ full-width half maximum (FWHM).
The Resolve instrument provides unprecedented spectral resolution for diffuse emission and extended objects representing a major advance over the CCD and gratings instruments on the Chandra X-ray Observatory and \xmmn.

\xrism\, observations of Cas~A were planned with various objectives that exploit the power of non-dispersive, high-resolution spectroscopy.  
The energy shifts of the bright lines/line complexes of Si, S, Ar, Ca, Fe, \& Ni were proposed to be used to reveal the velocity structure of the remnant to a level of precision that had not been possible with previous observations.  The high spectral resolution of Resolve allows the broadening of the bright lines to be measured to further constrain the three-dimensional structure and the conditions of the plasma, specifically the ion temperature.  The high quality data from Resolve also allow an exploration of the differences between the He-like emission (the He$\alpha$ triplets) and H-like emission (the Ly$\alpha$ lines) of Si and S by spectroscopically resolving different regions behind the shock front with different plasma conditions.  The high sensitivity of Resolve was proposed to be used to detect and characterize relatively weak emission lines from the odd-Z elements of P, Cl, and K to compare to SNe nucleosynthesis models in order to constrain the properties of the progenitor and the explosion.  Several papers have been prepared on these topics and three are included in this volume. 
The dynamics of Si and S making use of the H-like and He-like lines are presented in \citet{suzuki2025}. The velocity and broadening of the Fe-K complex is presented in \citet{bamba2025}. Maps of the Si and S velocities are presented in \citet{vink2025}.   A search for emission from the odd-Z elements of P, Cl, \& K is described in \citet{sato2025a} and has been submitted to another journal.  This paper serves as an introduction to the \xrism\, observations of Cas~A and discusses some technical aspects of the analysis.

All of these results depend upon the high spectral resolution and the calibration of the Resolve instrument.  They also depend upon the spectral models that are used to interpret the high-resolution data.  In this paper, we focus on the spectral models and how the results depend upon these models.
We compare results for two of the brightest regions in Cas~A that were observed on-axis with two plasma emission models, the {\texttt{APEC}} software/{\texttt{AtomDB}} database \citep{foster2012} and the {\texttt{SPEX}} software \citep{kaastra1996}.  Futhermore, we compare results with the versions of these software packages and associated atomic databases that were available before and after the launch of the \xrism\, mission. Some of the changes in the software and databases were motivated by the high-resolution spectra from Resolve, in particular the spectra from Cas~A.  We describe which of the model parameters are affected by these changes.
 
This paper is organized as follows.  In Section~\ref{sec:obs}, we describe the \xrism\, observations of Cas~A and the processing of the data relevant for this analysis. In Section~\ref{sec:analysis}, we present the image analysis, maps of the spatial-spectral mixing effect, and the spectral analysis with {\texttt{AtomDB}} and {\texttt{SPEX}}. In Section~\ref{sec:discuss} we discuss the results of the spectral fitting and the comparison of the spectral models.  All uncertainties presented in this paper are the $1.0\sigma$ uncertainties unless specified otherwise.

\begin{figure*}
\begin{center}
\includegraphics[trim ={0mm 0mm 0mm 0mm, clip}, width=7.5cm,angle=0]{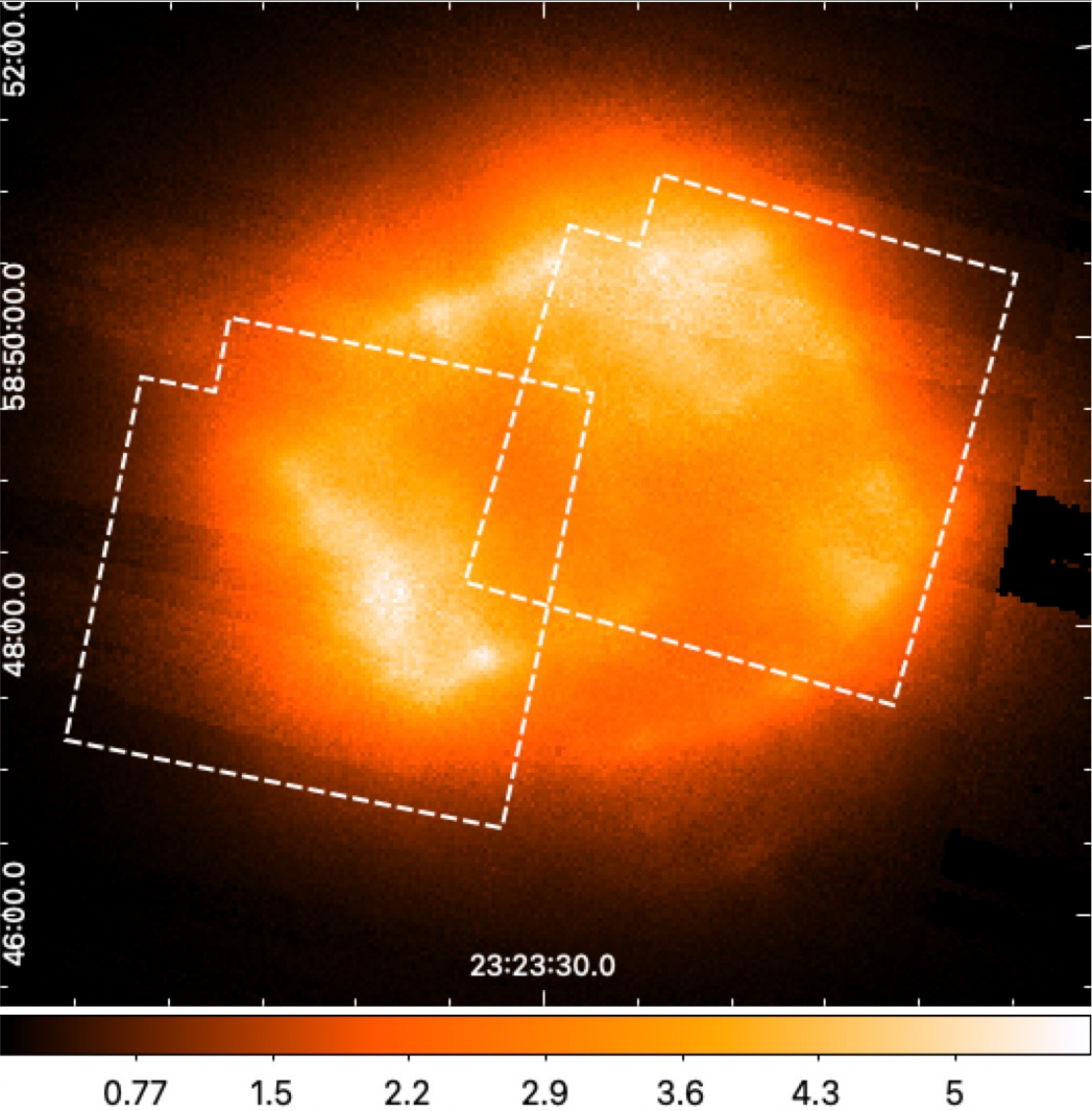}
\hspace{0.2cm}
\includegraphics[trim ={0mm 0mm 0mm 0mm, clip}, width=7.5cm,angle=0]{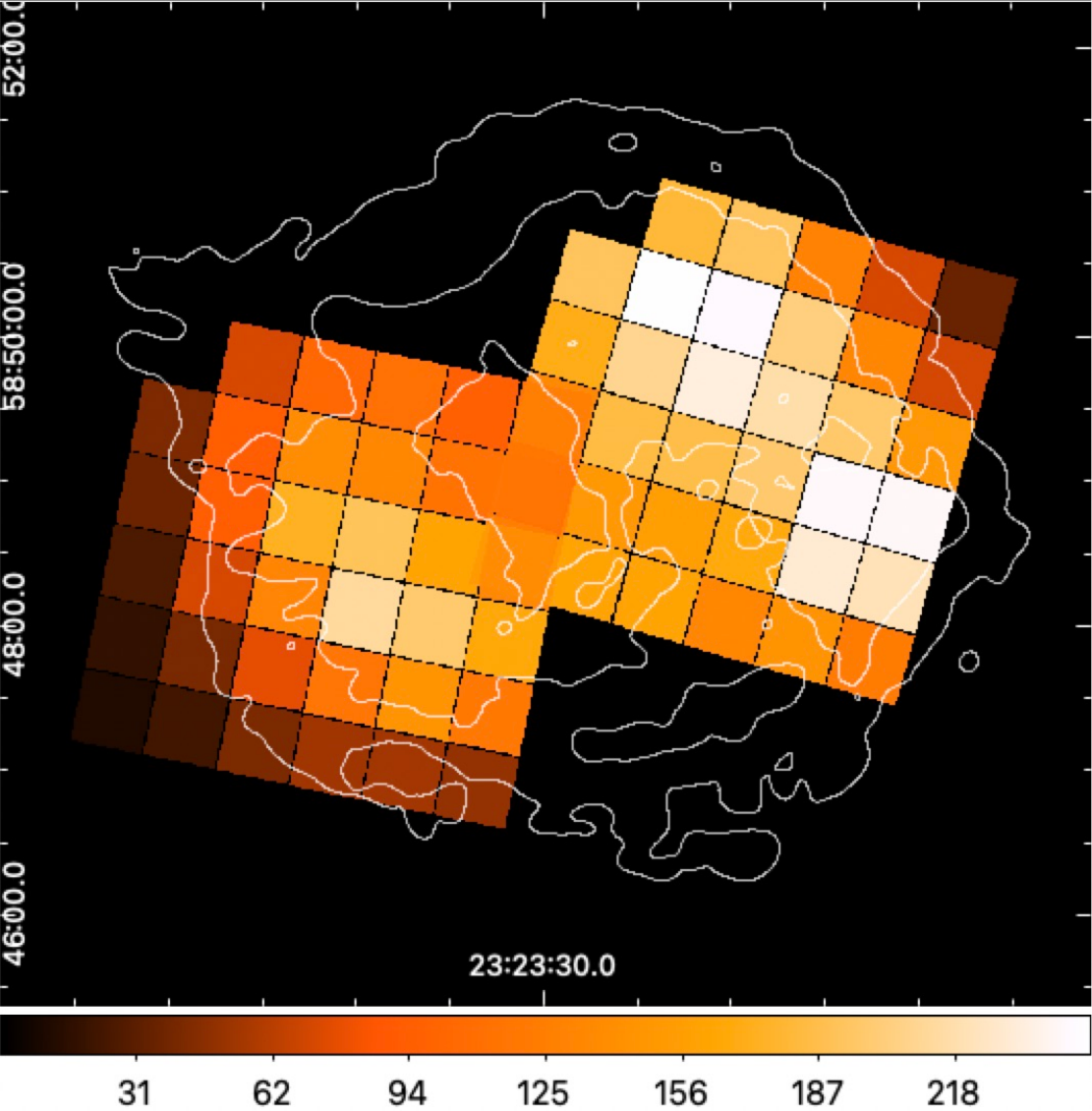}
\end{center}
\caption{(LEFT) Xtend counts image from the SE \& NW pointings in the 1.5-10.0~keV band. The Resolve FOVs are indicated by the dashed white squares. The data have been exposure corrected and the units are counts/ks in a pixel. 
(RIGHT) Resolve counts images from the SE and NW pointings in the 1.5-10.0 keV band. The individual pixels in the array are indicated by the small black squares. The data have been exposure corrected and the units are counts/ks in a pixel.
The contours show Cas A as derived from a broadband image of a \chandra\ observation (ObsID 4638).
{Alt text: Two color maps, one on the left and one on the right, that show the counts/ks in each pixel in the Xtend detector (LEFT) and the Resolve detector (RIGHT). A color bar at the bottom shows the relationship between color and counts/ks.} 
}
\label{fig:images}
\end{figure*}

\section{Observations and Data Reduction}
\label{sec:obs}

\begin{table}
  \tbl{\xrism\, Observation Log of Cas~A }{%
   \begin{tabular}{cccccc}
      \hline
      OBSID & RA       & Dec     & Roll   & Start   & Exposure\footnotemark[$*$] \\
            & (J2000)  & (J2000) & (deg)  &  Date   & (ks) \\
      \hline
000129000  & 350.930  & +58.808  & 258.52  & 2023345.243  & 181.3 \\
000130000  & 350.862  & +58.822  & 254.06  & 2023348.584  & 165.7  \\
      \hline
    \end{tabular}}
    \label{tab:obs}
\begin{tabnote}
\footnotemark[$*$] Accepted Resolve exposure after applying standard filtering criteria.  \\ 
\end{tabnote}
\end{table}

\xrism\ observed Cas~A twice during the Performance Verification (PV) phase, from December 11 to 14, 2023 targeting the southeast region (``SE" hereafter, ObsID 000129000) and from December 14 to 17, 2023 targeting  the northwest region (``NW" hereafter, ObsID 000130000). The nominal aim point, roll angle, and accepted Resolve exposure times for both observations are listed in Table~\ref{tab:obs}. The observations were made with the Resolve dewar gate valve closed, which restricted the bandpass to energies above $\sim 1.5$~keV.

\begin{figure}
\begin{center}
  \includegraphics[clip, trim =20mm 0mm 30mm 0mm, width=8.5cm,angle=0]{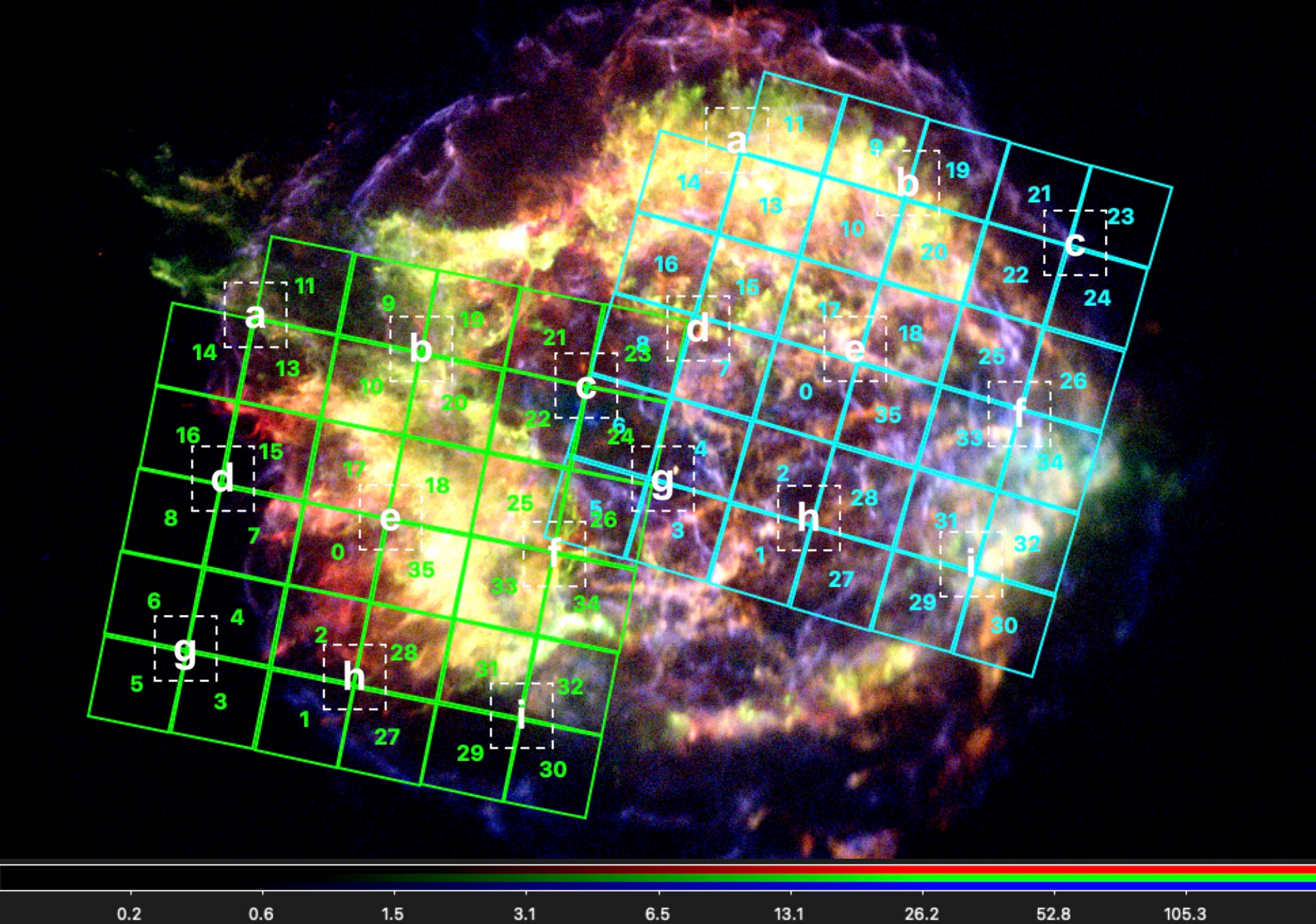} 
\end{center}
\caption{Resolve pixel arrays overplotted on the \chandra\, true color image with the individual pixels numbered 0-35 and the super-pixels labelled a-i.
{Alt text:  A color image that shows Resolve arrays with individual pixels represented as squares, overplotted on the \chandra\, true-color image. A color bar at the bottom shows the relationship between color and intensity in the red, green, \& blue bands.} 
}
\label{fig:resolve_pixels}
\end{figure}

The observation data were processed using the pre-pipeline software version \texttt{004\_001.15Oct2023\_Build7.011} and the processing (pipeline) script \texttt{03.00.011.008}. 
The Resolve detector gain was calibrated on-orbit using 55 fiducial gain measurements (29 for SE and 26 for NW) of $^{55}$Fe radioactive sources during Earth occultation, achieving an energy resolution of 4.55 $\pm$ 0.02 eV (FWHM) and 4.53 $\pm$ 0.02 eV (FWHM) at 5.9 keV for SE and NW respectively \citep{porter2024}. The energy-scale error using the constantly illuminated calibration pixel was 0.23 eV and 0.31 eV for the SE and NW, respectively, which is comparable to other \xrism\ observations. 
In this early observation, the fiducial gain measurements were made more frequently enabling precise gain tracking of all pixels. Thus we include pixel $\#27$ (see Figure~\ref{fig:resolve_pixels} for the pixel numbering scheme), which has been shown to exhibit erratic gain jumps, and perform the analysis using the full array for both observations.

The Resolve data for both observations were reprocessed with the HEASoft 6.34 software package, applying calibration from \xrism\ CalDB 9 (Version 20240815). The default screening as described in the \xrism\ team’s analysis of N132D (XRISM Collaboration 2024) was applied, resulting in clean exposure times of 181.3 and 165.7 ks for SE and NW, respectively. 
We do not apply any additional good time interval (GTI) screening based on the pointing stability of the satellite.
Only high-resolution primary (‘Hp’ or \texttt{ITYPE}=0) grade events were included for the following analysis, as they account for more than $96\%$ of the 2–10 keV events in each observation.
A redistribution matrix file (RMF) was generated in extra large (XL) size in split-rmf format by the \texttt{rslmkrmf} task using the cleaned event files. 
This RMF incorporates: Gaussian core, low-energy exponential tails, Si fluorescence lines, escape peaks and electron loss continuum. 
The anciliary response file (ARF) was constructed with \texttt{xaarfgen} task, using an exposure-corrected Chandra image in the 0.5-5 keV band as an input model for source flux.
A non-X-ray background (NXB) spectrum was modeled from a database\footnote{ https://heasarc.gsfc.nasa.gov/docs/xrism/analysis/nxb/index.html} of Resolve night-Earth data using \texttt{rslnxbgen} while the sky background was ignored. 
The spectra were binned using \texttt{SPEX v3.08.01} software package applying the optimal binning method of \cite{kaastra1996}, where each bin had at least 1 count to fit the spectra using the C-statistic, for the fits performed in \texttt{SPEX} and were binned with the \texttt{ftool ftgtoupha} with the optimal binning option for the fits performed in \texttt{XSPEC}.  

\section{Analysis and Results}
\label{sec:analysis}

\subsection{Image Analysis}

\subsubsection{Xtend and Resolve Images}
\label{sec:image_analysis}

Figure~\ref{fig:images} shows the Xtend image of Cas~A in the 1.5-10.0 keV band from the SE and NW observations in the left panel.  The Xtend data cover all of Cas~A and the surrounding regions due to its large FOV. This image has been extracted from those data to show only the region around Cas~A. The Xtend image shows the familiar partial shell-like structure of Cas~A known from the \chandra, \xmm, and earlier observations with the bulk of the emission arising from the shock-heated ejecta. The two Resolve FOVs are overplotted on the Xtend image as shown by the dashed white squares. 
The Resolve images are shown in the right panel of Figure~\ref{fig:images}.  Both images present the counts/ks in a pixel in the 1.5-10.0~keV band,  indicating the statistical quality of the data for spectroscopy.
Note that the calibration pixel (pixel $\#12$) in the upper left corner is omitted from the white square since it contains no data from the remnant.
Figure~\ref{fig:resolve_pixels} shows the pixel numbering scheme for the two Resolve pointings overplotted on the \chandra\, true-color image and the definition of the super-pixels ``a-g'' that will be used in section~\ref{sec:on-axis_spectra}.  The
super-pixels are simply $2\times2$~pixel regions, except for super-pixel ``a'' which is missing one of the four pixels, pixel $\#12$.

The SE pointing was positioned to maximize coverage of the previously known Si-rich and Fe-rich regions.  
The NW pointing was positioned to cover most of the arc of emission in the West, part of the bright arc in the North (which is also rich in Si and Fe), and the region towards the center that is apparently interacting with the shell of circumstellar medium (CSM) material.  
The Xtend image shows the parts of the remnant that are outside of the Resolve FOV. Regions that are outside of, but close to, the Resolve FOV will contribute significantly to the observed counts as discussed in detail in the next section~\ref{sec:ssm}.  The Xtend and Resolve images both show that spatial differences can be observed even with the moderate angular resolution of \xrism.

\subsubsection{Spatial-Spectral Mixing}
\label{sec:ssm}

The in-orbit on-axis Point Spread Function (PSF) of \xrism\ is 1.3’ half-power diameter (HPD) \citep{hayashi2024} which is relatively large compared to the Resolve pixel size of 0.5’. This leads to photons from neighboring regions cross-contaminating the Resolve pixel spectra. Given that this happens at all energies and a Resolve spectrum is associated with each spatial pixel, this mixing affects the Resolve spectrum both spatially and spectrally, thus referred to as spatial-spectral mixing (SSM). In addition to the effects of PSF, small wobbles in the \xrism\ attitude can also result in SSM. As these can affect the physical parameter mapping of extended sources, we want to estimate their severity using simulated data.

\begin{figure*}
 \begin{center}
   \includegraphics[trim ={0mm 0mm 0mm 0mm, clip}, width=9.2cm,angle=0]{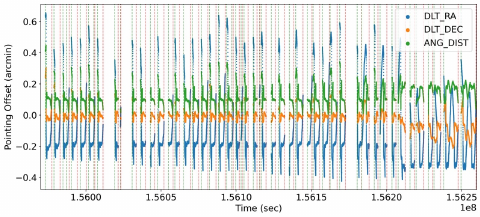} 
   \includegraphics[trim ={0mm 0mm 0mm 0mm, clip}, width=4.2cm,angle=0]{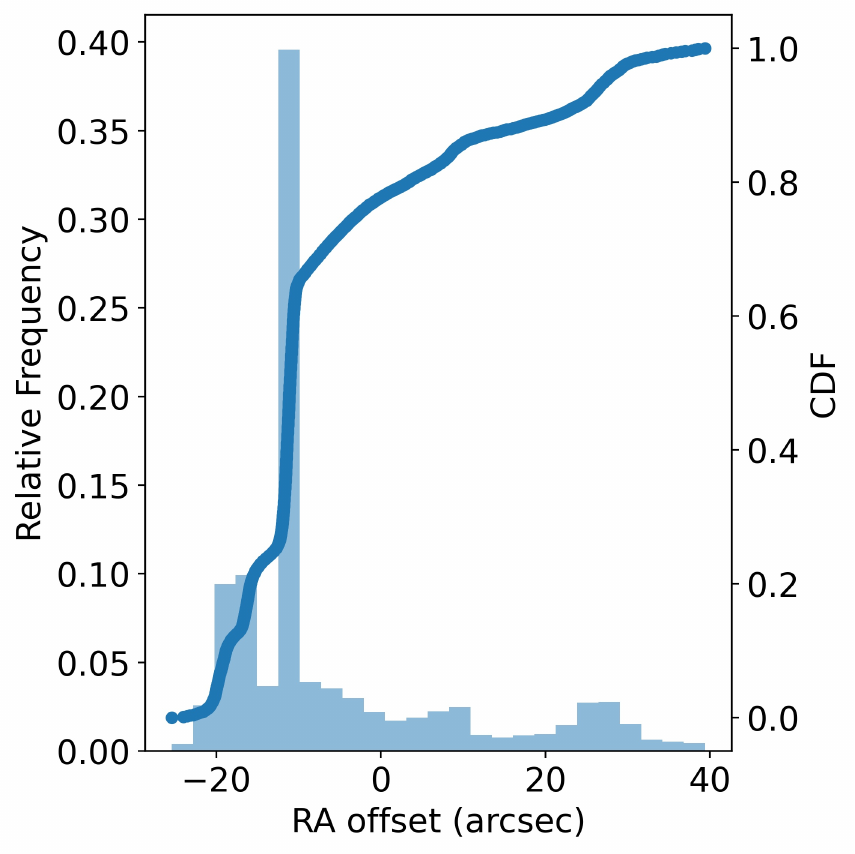}
   \includegraphics[trim ={0mm 0mm 0mm 0mm, clip}, width=4.2cm,angle=0]{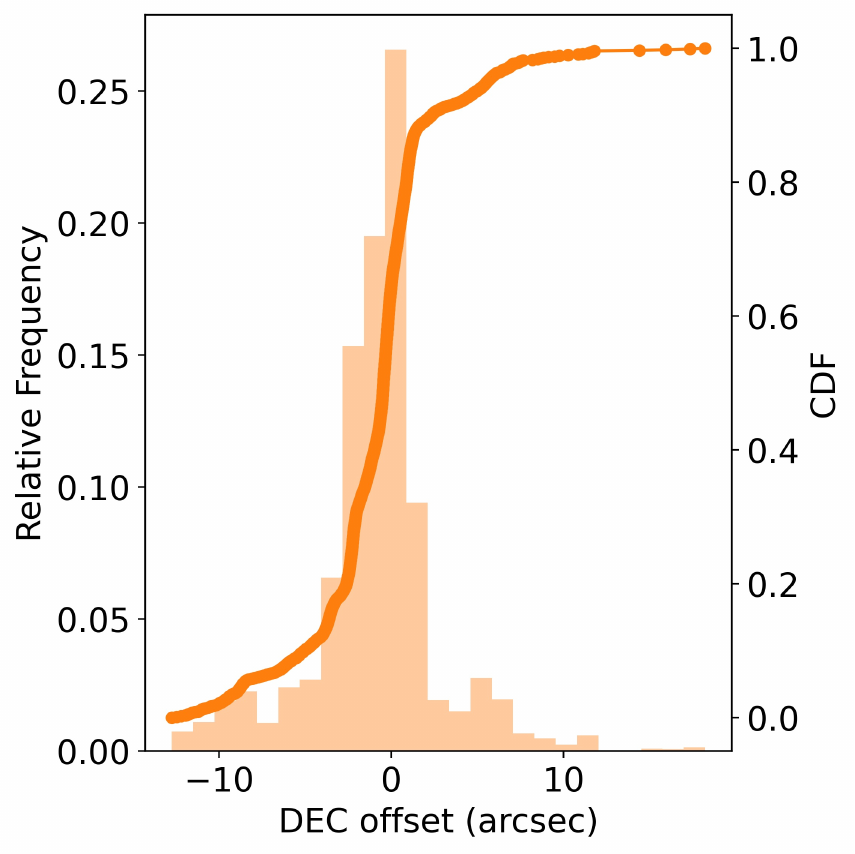} 
 \end{center}
\caption{The \xrism\ pointing stability during the SE observation of Cas~A. The left panel shows the pointing offset after applying the GTI correction. The blue (DLT\_RA) and orange (DLT\_DEC) lines show the offset in RA and DEC, respectively, and ANG\_DIST shows the net angular distance from the nominal pointing (RA\_NOM, DEC\_NOM).  The green and red dotted lines show the GTI start and stop time, respectively. The two panels on the right show the distribution of offset in RA and DEC, and the corresponding cumulative distribution functions (CDF) which are used in a Monte Carlo approach to convolve the photon sky locations to account for attitude effects.
{Alt text: The left most panel shows line graphs corresponding to the pointing offset in RA, DEC and total angular distance as a function of time. The two panels on the right, show histogram distribution and CDF of the offset in RA and DEC in units of arcsec.} 
}
 \label{fig:SE_attitude}
\end{figure*}

We implemented a Monte Carlo-based algorithm to convolve the data from X-ray missions with better spatial resolution (e.g. \chandra\, or \xmm) to the \xrism\ response. We take an event file and perform Monte Carlo simulations on the sky location of each photon by drawing from the cumulative distribution functions of the PSF and the attitude profiles. The \xrism\ PSF varies with photon energy and off-axis angle, and has a complex shape due to mirror support structures, imperfect alignments, and non-ideal foil surfaces. However, for our analysis, we used the on-axis PSF from the encirled energy function (EEF) for 4.5 keV photons, thus assuming an energy-independent and azimuthally symmetric PSF. To account for attitude effects, we used the distribution of aim-point offsets (separately for RA and DEC) after applying the good time interval (GTI) correction. 
The early PV phase observations of Cas~A were performed before the ground software update that refined the Earth occultation prediction for the
star trackers under specific attitude and orbit conditions. During these observations, the attitude and orbit control system relied on the
inertial reference units for longer periods. The resulting attitude variations were therefore larger than those in later observations, and
no other scientific observations were affected by this issue.
Figure~\ref{fig:SE_attitude} shows the GTI corrected pointing offset for the the SE observation and the corresponding cumulative distributions that are used to perform the Monte Carlo simulation.

\begin{figure*}
 \begin{center}
   \includegraphics[trim ={0mm 0mm 0mm 0mm, clip}, width=8.0cm,angle=0]{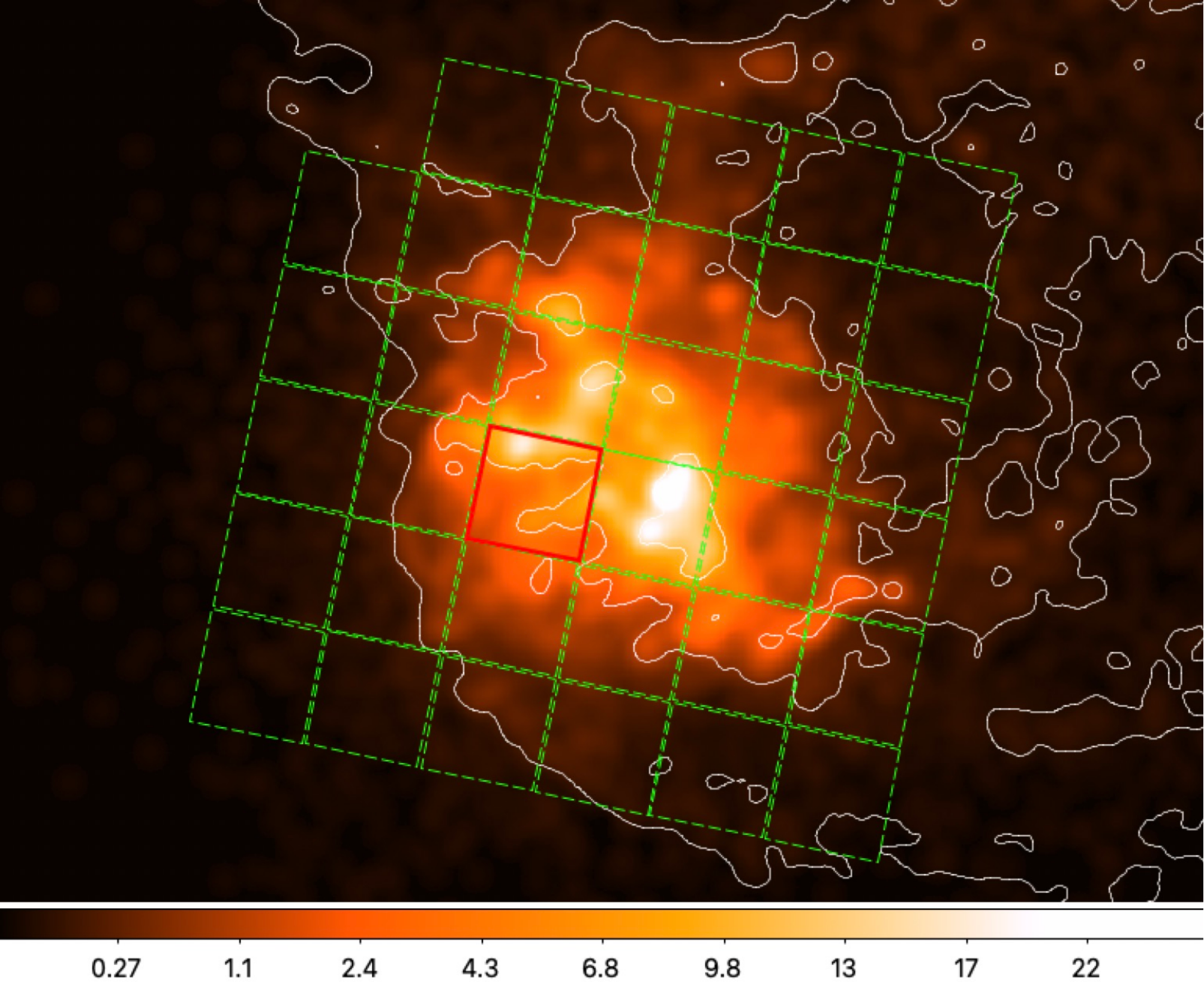} 
   \includegraphics[trim ={0mm 0mm 0mm 0mm, clip}, width=8.0cm,angle=0]{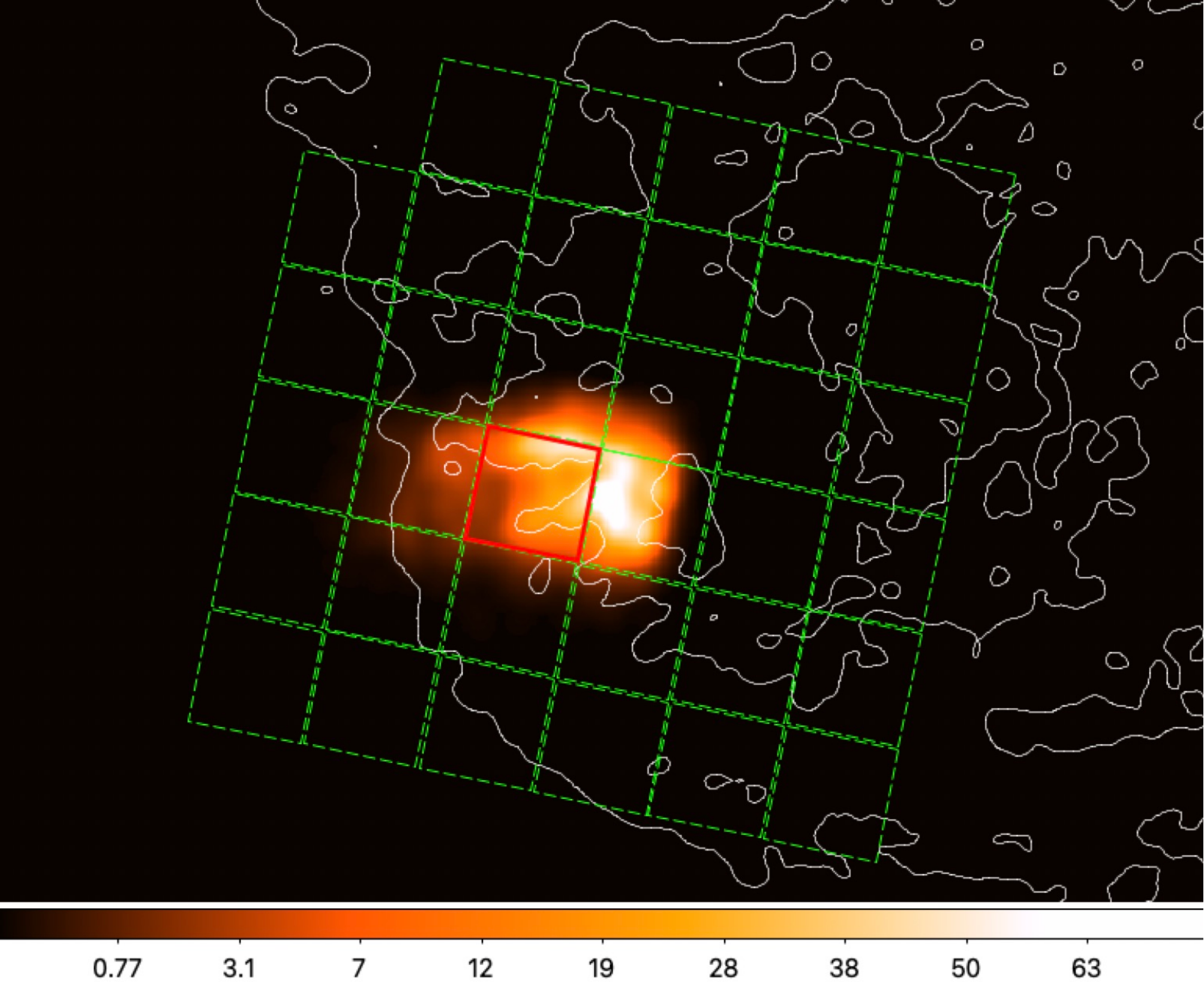}
 \end{center}
\caption{The sky location of all the photons in the energy range 1.8 to 7 keV that are detected at the Resolve pixel $\#0$ (marked with red border). The dashed green boxes outline the Resolve pixel array for the SE observation. The left panel shows the effects due to the relatively large PSF of \xrism\ and the right panel shows the blending due to the attitude effects alone. The images are smoothed with a gaussian kernel to represent the small variation over different Monte Carlo simulation runs. The white contours are from \chandra\ broadband image. The photon counts are shown with a square root scale.
{Alt text: Two sky maps showing the distribution of photons that hit the Resolve pixel $\#0$ due to PSF and attitude effects.} 
}
 \label{fig:SE_trackback_maps}
\end{figure*}

\begin{figure*}
 \begin{center}
   \includegraphics[trim ={0mm 0mm 0mm 0mm, clip}, width=8.7cm,angle=0]{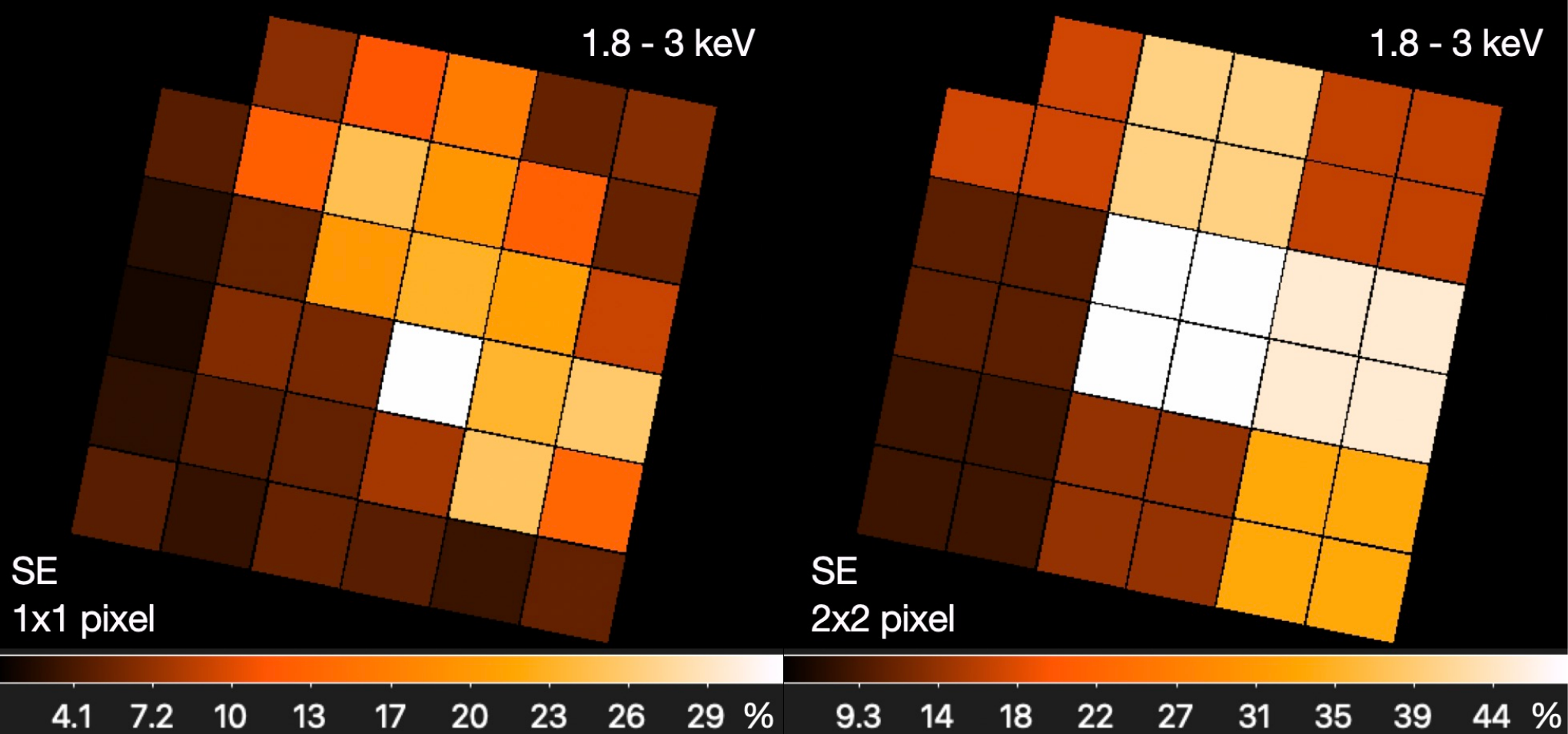} 
   \includegraphics[trim ={0mm 0mm 0mm 0mm, clip}, width=8.6cm,angle=0]{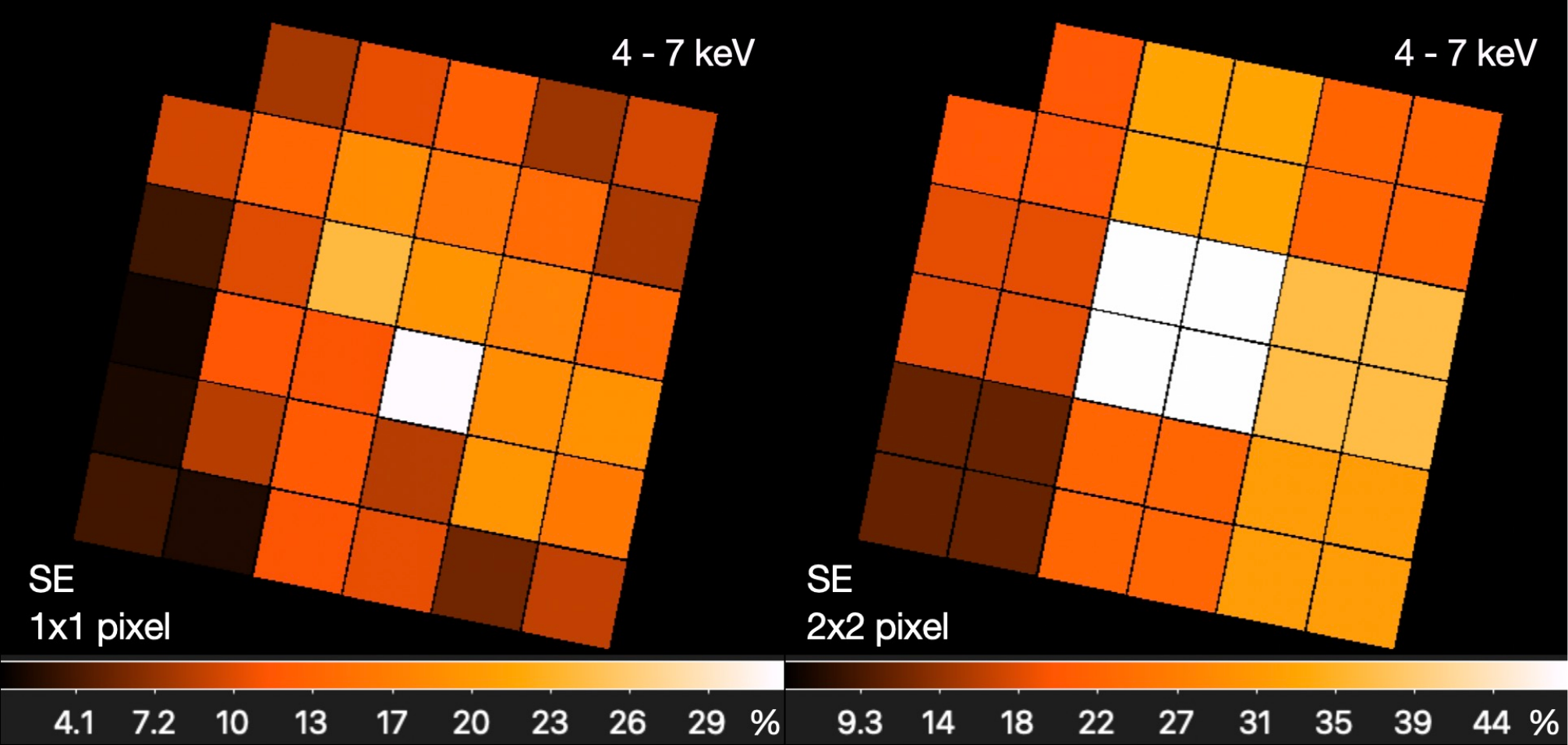}
 \end{center}
\caption{The percentage of photons that are detected at a Resolve pixel which actually originate from the corresponding sky region. These purity maps are shown for the SE observation for per Resolve pixel and $2\times2$ pixel regions. The left two panels show purity percentage of photons in the energy range 1.8 to 3 keV and the the right two panels for the energy range 4 to 7 keV.
{Alt text: Four heatmaps of Resolve pixel array showing the percentage of detected photons that truly originate from the sky region corresponding to the pixels.} 
 }
 \label{fig:SE_purity_maps}
\end{figure*}

\begin{figure*}
 \begin{center}
   \includegraphics[trim ={0mm 0mm 0mm 0mm, clip}, width=7cm,angle=0]{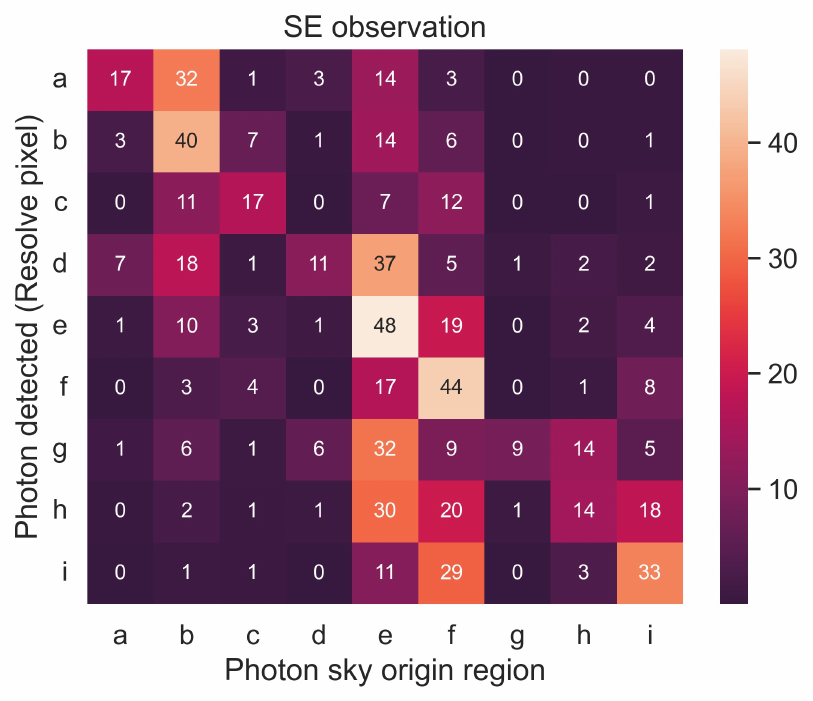} 
   \hspace{0.5cm}
   \includegraphics[trim ={0mm 0mm 0mm 0mm, clip}, width=7cm,angle=0]{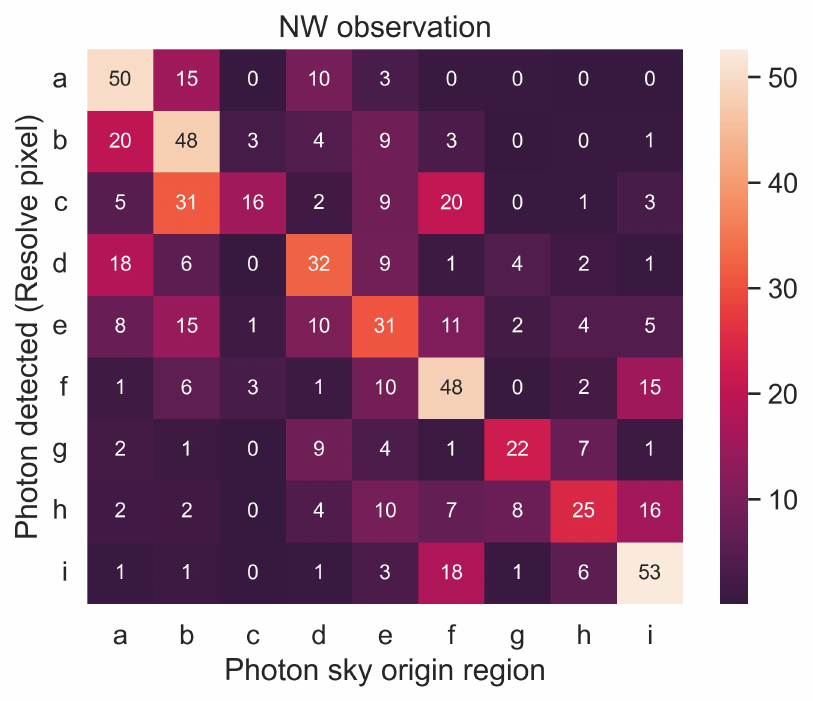}
 \end{center}
\caption{The estimated photon (in the energy range 1.8 to 7 keV) cross contamination percentage  per $2\times2$ Resolve pixel regions. This shows the extent of internal SSM for the SE and NW observation by accounting for both PSF and attitude effects. The regions labeled \texttt{a} to \texttt{i} are marked in Figure~\ref{fig:int_spectrum}. The colorbar is in units of percentage.
{Alt text: Two matrices showing the photon contribution percentage between the $2\times2$ pixel regions for the two Cas~A observation.} 
 }
 \label{fig:SE_blending_fractions}
\end{figure*}

We used the 2009 \chandra\ observation of Cas~A (ObsID 4638) with an exposure time of 160 ks and a total photon count of $\sim50$ million. We only considered photons in the energy range 1.8 to 7 keV, as this bandpass overlaps with that of Resolve, which accounts to $\sim23$ million photons and we ignored background. The \chandra\ sky coordinates of a photon are considered as its "true" sky location, and then an offset (separately for RA and DEC) is added to account for the SSM. Since we know the sky location of the photon before and after the convolution, we can precisely track back the photons from the detector plane to their origin on the sky. Figure~\ref{fig:SE_trackback_maps} shows the "true" sky location of the photons that are eventually detected at Resolve pixel $\#0$ (marked with a red border), the left panel accounts for only PSF effects, and the right panel shows attitude effects for SE observation alone. We estimate that $\sim80\%$ of the photons originate from outside the sky region covered by Resolve pixel $\#0$ due to just PSF effects and $\sim50\%$ due to attitude effects alone. When both effects are considered, only $\sim6\%$ of the photons come from the sky region covered by pixel $\#0$, we refer to this as the "purity percentage". Figure~\ref{fig:SE_purity_maps} shows maps of the purity percentage per pixel and per $2\times2$ pixel region for the SE observation. The two leftmost images show the purity percentage maps for photons in the energy range 1.8 to 3 keV, and the two rightmost images show maps for the 4 to 7 keV range. In addition, we also estimate the contribution from other pixel regions. For example, in the case of Resolve pixel $\#0$ the majority of photons ($\sim27\%$) originate from the sky region corresponding to Resolve pixel $\#35$. This cross-contamination between pixels, i.e. internal SSM,  can be used to account for SSM while spectral fitting. Figure~\ref{fig:SE_blending_fractions} shows the cross-contamination percentages in the SE and NW observations for $2\times2$ Resolve pixel regions (the so-called ``super pixels''), which are labeled \texttt{a} to \texttt{i} as indicated in Figure~\ref{fig:resolve_pixels}. The diagonal of the matrix in Figure~\ref{fig:SE_blending_fractions} is the same as a purity percentage map. We find that the highest purity percentage for Cas~A observations for individual pixels is $\sim30\%$ and increases to $\sim50\%$ when using $2\times2$ pixels regions. We also evaluate the External SSM, i.e. the contribution of photons from outside the Resolve field of view, as $\sim24\%$ and $\sim22\%$ for the SE and NW observations, respectively.

\subsection{Spectral Analysis}

\subsubsection{Integrated Spectra from Each Observation}
\label{sec:integrated_spectra}

\begin{figure}
\begin{center}
 \includegraphics[trim ={0mm 10mm 0mm 0mm, clip=true}, width=8.0cm,angle=0]{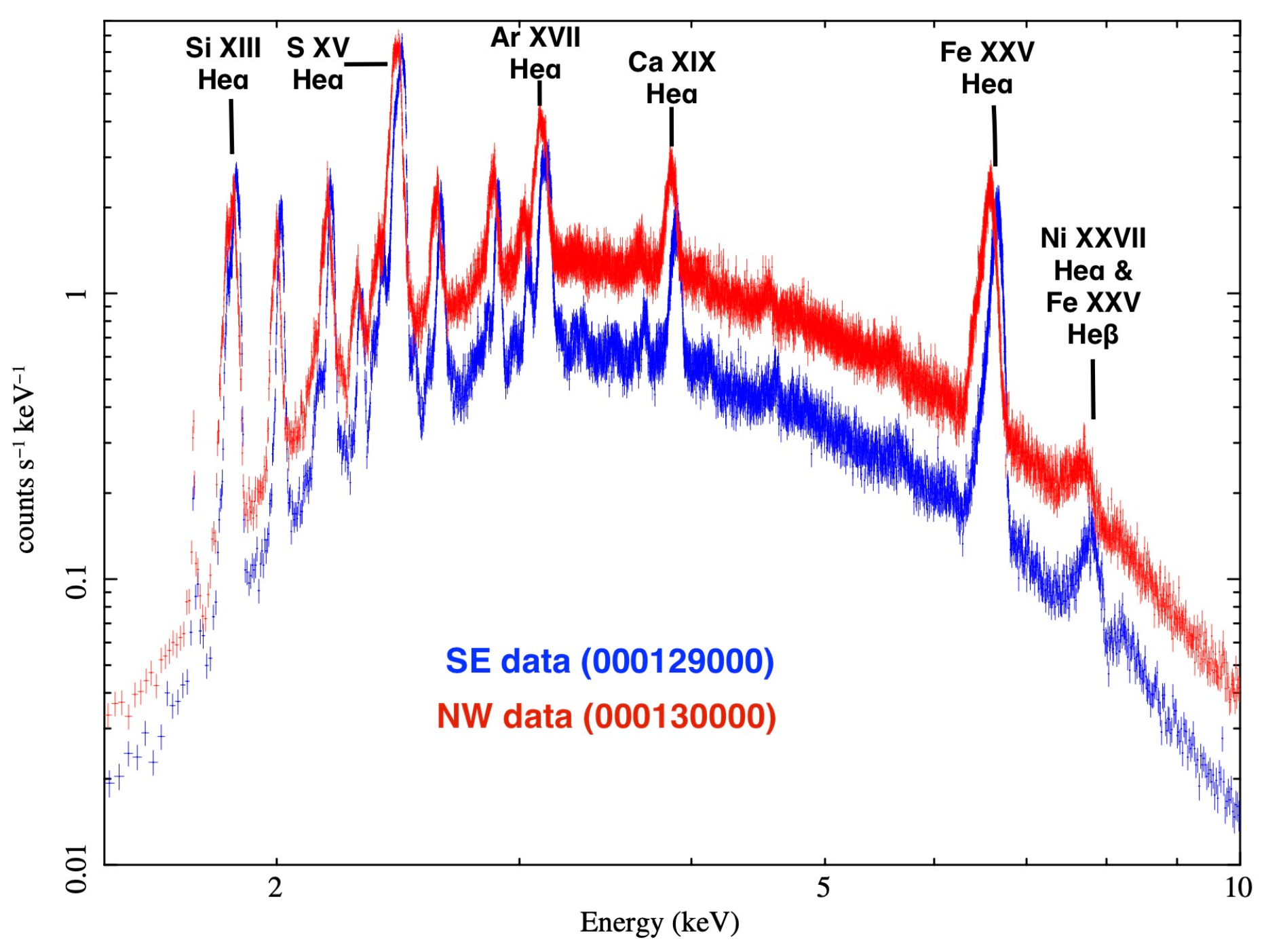}  
\end{center}
\caption{Resolve spectra from the entire array from the SE pointing (blue) and the NW pointing (red).  The prominent line complexes of Si, S, Ar, Ca, \& Fe are labeled. The redshift/blueshift between the observations is apparent.
{Alt text: A line graph showing two spectra, one from the entire array for the SE observation and one for the NW observation. The horizontal axis shows the energy in keV and the vertical shows the ${\mathrm{count~s^{-1}~keV^{-1}}}$.} 
}
\label{fig:int_spectrum}
\end{figure}

Spectra were extracted from all pixels, except for the calibration pixel $\#12$, to create an integrated spectrum for each observation (SE \& NW)  using the  HEASoft tool \texttt{XSELECT}.  These two spectra are plotted in the left panel in Figure~\ref{fig:int_spectrum}, the SE spectrum in blue and the NW spectrum in red.  These spectra are dominated by the bright line complexes of Si, S, Ar, Ca, Fe, \& Ni, with many weaker lines evident in the spectra.
It is important to note that these spectra are derived from a mixture of multiple regions within the remnant, each with its own plasma conditions (temperature, ionization timescale, abundances, etc.) and velocity structure as described in the previous section ~\ref{sec:ssm}. 
This mixing of emission from different regions broadens the lines significantly beyond the resolution of the Resolve calorimeter complicating efforts to measure the intrinsic thermal broadening.
Even with this large degree of mixing and the associated broadening, it is clear that there is a systematic energy shift between these two spectra.  The SE spectra are blue-shifted with respect to the NW spectrum.  This is the same blue-shift/red-shift pattern from SE to NW in the remnant that has been observed in earlier observations of Cas~A \citep{markert1983,holt1994,willingale2002,lazendic2006}.  

\begin{figure}
\begin{center}
 \includegraphics[trim ={0mm 10mm 0mm 0mm, clip=true}, width=8.5cm,angle=0]{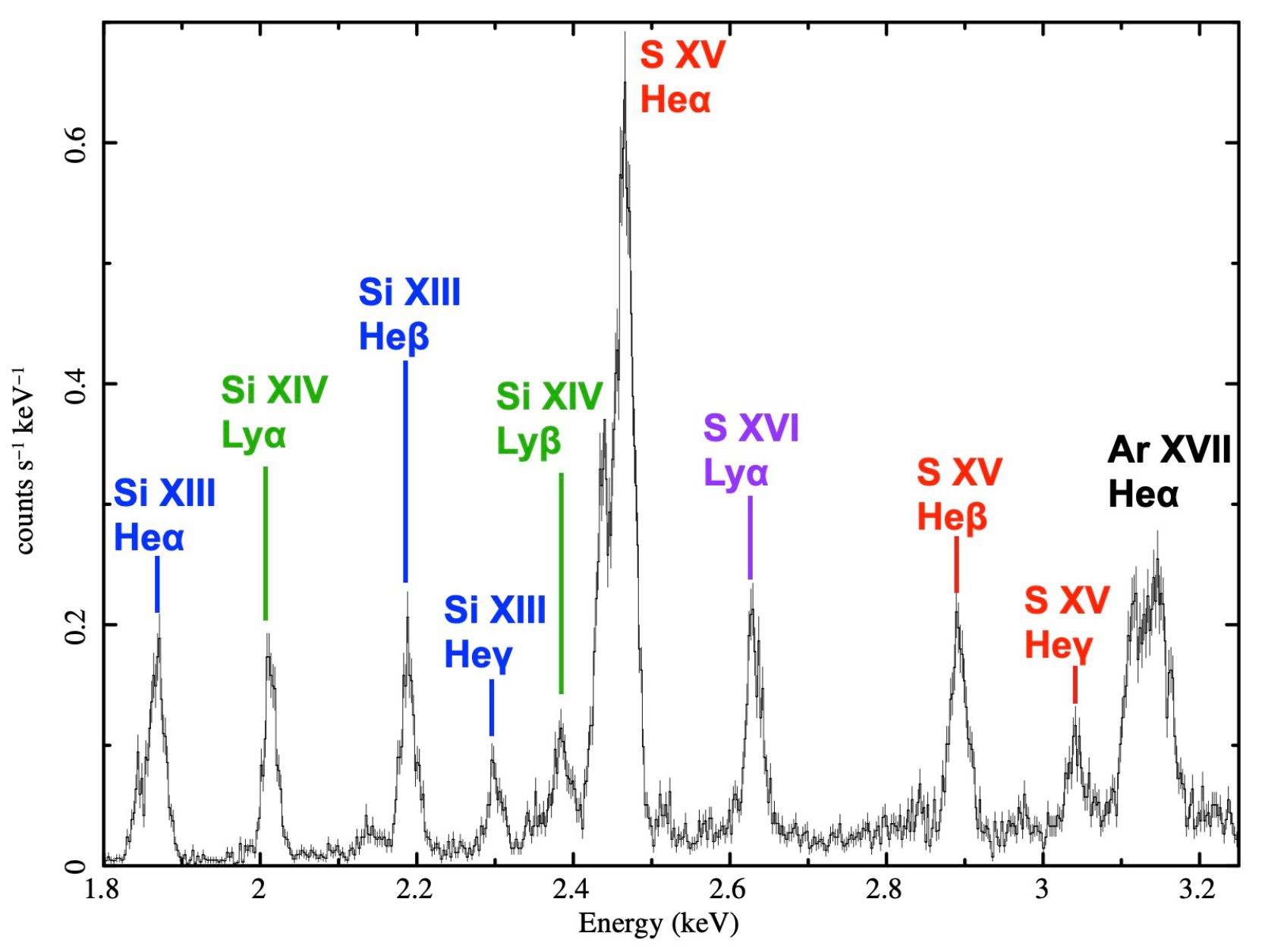} 
\end{center}
\caption{Resolve spectra from pixel ${\mathrm {\#35}}$ in the SE pointing.  The line complexes of Si~XIII, Si~XIV, S~XV, S~XVI, \& Ar~XVII are labelled. 
{Alt text: A line graph on the left showing one spectrum with the bright line complexes labeled. The horizontal axis shows the energy in keV and the vertical axis shows the data in units of ${\mathrm{count~s^{-1}~keV^{-1}}}$.}

}
\label{fig:SiS_spectrum}
\end{figure}

Spectra extracted from single pixels or small groups of adjacent pixels have significantly less broadening than the spectra extracted from the entire array. A map of the Resolve pixels over-plotted on the \chandra\, image of Cas~A is shown in Figure~\ref{fig:resolve_pixels}.  The individual pixels are numbered in this image and groups of $2\times2$ ``super pixels'' are labelled a-i.  To more clearly demonstrate the spectral resolving power of the calorimeter, we extracted a spectrum from a single pixel, pixel $\#35$ in the SE pointing which has the highest purity as shown in Figure~\ref{fig:SE_purity_maps}.  Figure~\ref{fig:SiS_spectrum} displays this spectrum showing that the Si~{\footnotesize{XIII}} and 
S~{\footnotesize{XV}} He$\alpha$, He$\beta$, He$\gamma$, the Si~{\footnotesize{XIV}} Ly$\alpha$ and Ly$\beta$, the S~{\footnotesize{XVI}}~Ly$\alpha$, and the Ar~{\footnotesize{XVII}}~He$\alpha$ lines/line complexes are cleanly resolved.  There is also a S~{\footnotesize{XVI}}~Ly$\beta$ line which is blended with the forbidden line of the Ar~{\footnotesize{XVII}}~He$\alpha$ triplet since the energies of these lines only differ by $\sim2$~eV. This spectrum demonstrates the power of the Resolve calorimeter to perform high resolution spectroscopy in this energy range even with the \xrism\, Gate Valve closed.
The broadening of the lines in the spectrum from a single pixel is significantly less than that in the spectra from the entire array.  This can be seen in Figure~\ref{fig:SiS_spectrum} as the forbidden and resonance lines of the S~{\footnotesize{XV}}~He$\alpha$ triplet are resolved from each other.  The three companion papers in this issue, \cite{suzuki2025, bamba2025, vink2025},  measure the energy shifts and the broadening of these lines/line complexes to characterize the plasma conditions and the three-dimensional structure of Cas~A taking advantage of the high spectral resolution and precise gain calibration of the Resolve calorimeter.

\subsubsection{Spectra from Bright, On-axis Regions}
\label{sec:on-axis_spectra}

\begin{figure*}
\begin{center}
\includegraphics[clip, trim =0mm 5mm 50mm 25mm, width=8.5cm,angle=0]
{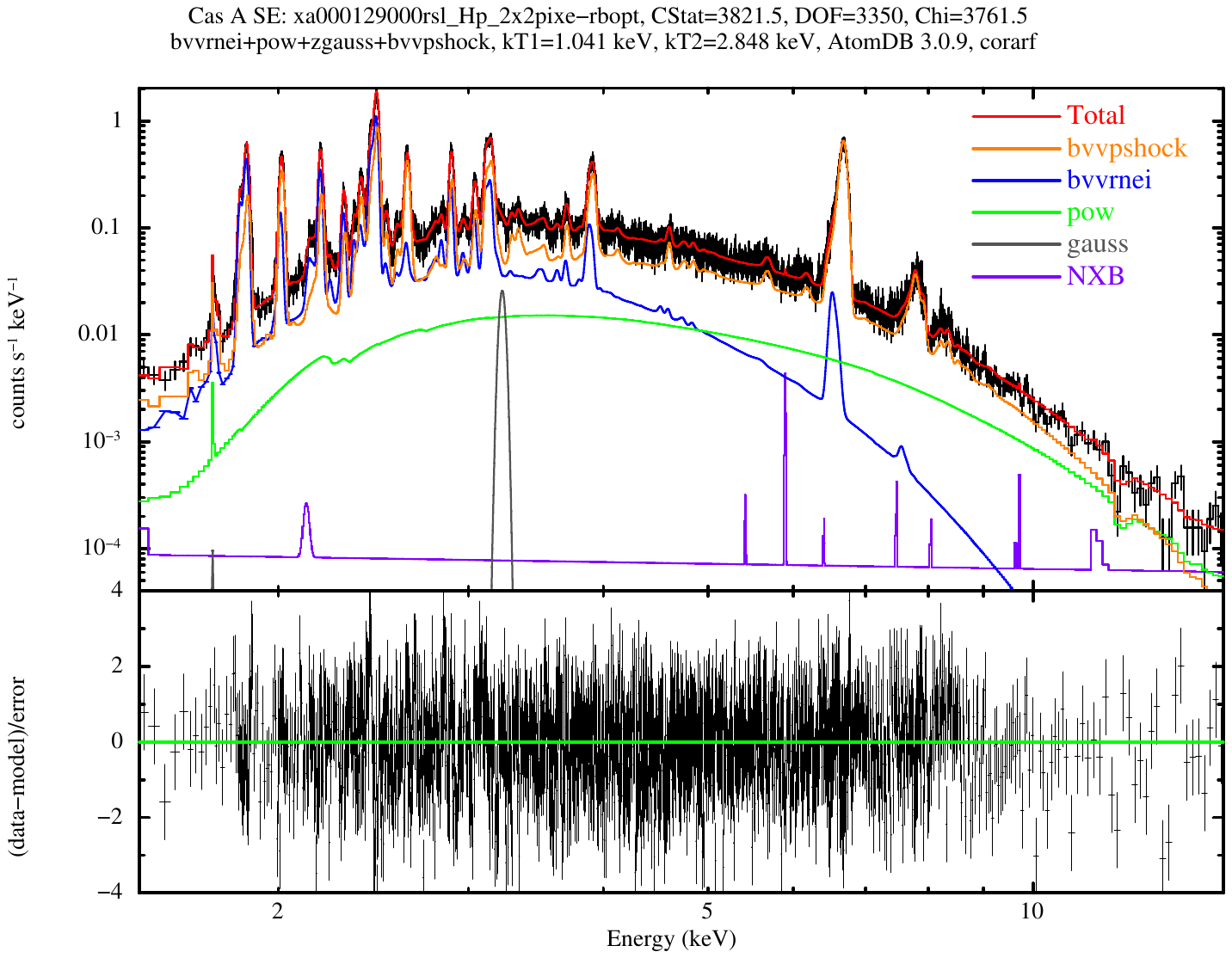} 
\includegraphics[clip, trim =0mm 5mm 50mm 25mm, width=8.5cm,angle=0] 
{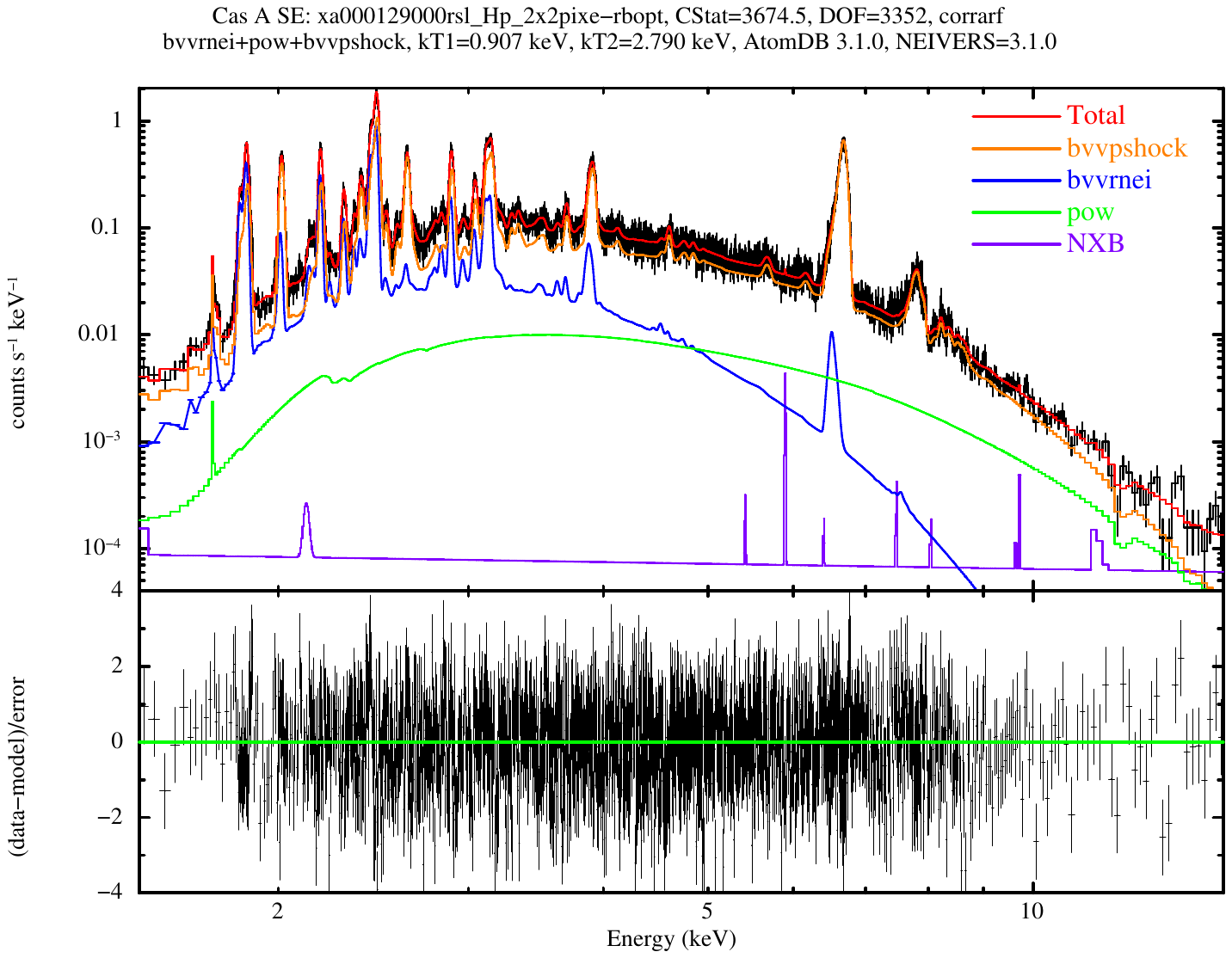}
\end{center}
\caption{LEFT: Resolve spectra from the on-axis point, so-called super-pixel ``e'' for the SE pointing in the 1.5-15.0 keV range fit with XSPEC and AtomDB~3.0.9. RIGHT: Resolve spectra from the on-axis point, super-pixel ``e'' for the SE pointing in the 1.5-15.0 keV range fit with XSPEC and AtomDB~3.1.0.  The data are represented by the black data points and have been rebinned for display purposes only. The upper red curve is the total model, the orange curve is the high~kT bvvpshock component, the blue curve is the low~kT bvvrnei component, the green curve is the power-law component, the gray line at $\sim3.2$~keV is the additional Gaussian component, and the purple curve is the NXB component. The residuals are plotted in the lower panel.
{Alt text: Two line graphs showing the same spectrum fit with different models, AtomDB 3.0.9 on the left and 3.1.0 on the right.  Additional lines show the model components separately. The lower panel shows the residuals for the fit. The plots use the same horizontal axis that shows the energy in keV.
The spectral plot has a vertical axis that shows the data and model in units of  ${\mathrm{count~s^{-1}~keV^{-1}}}$. The residual plot shows the residuals in units of data minus the model divided by the error.}
}
\label{fig:se_atomdb3.0.9_3.1.0}
\end{figure*}

\begin{figure*}
\begin{center}
 \includegraphics[clip, trim =0mm 0mm 0mm 15mm, width=8.5cm,angle=0]{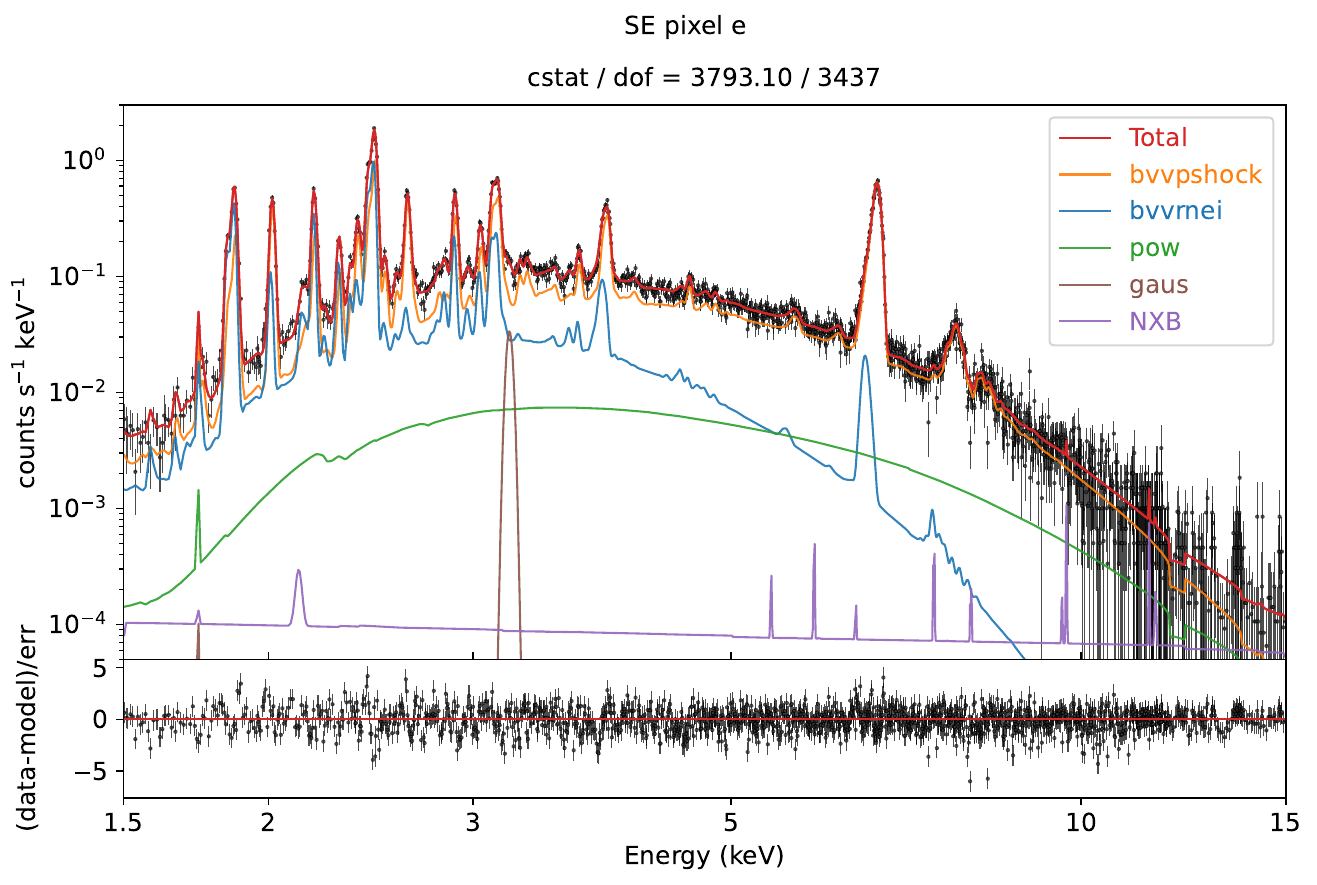} 
\includegraphics[clip, trim = 0mm 0mm 0mm 15mm, width=8.5cm,angle=0]{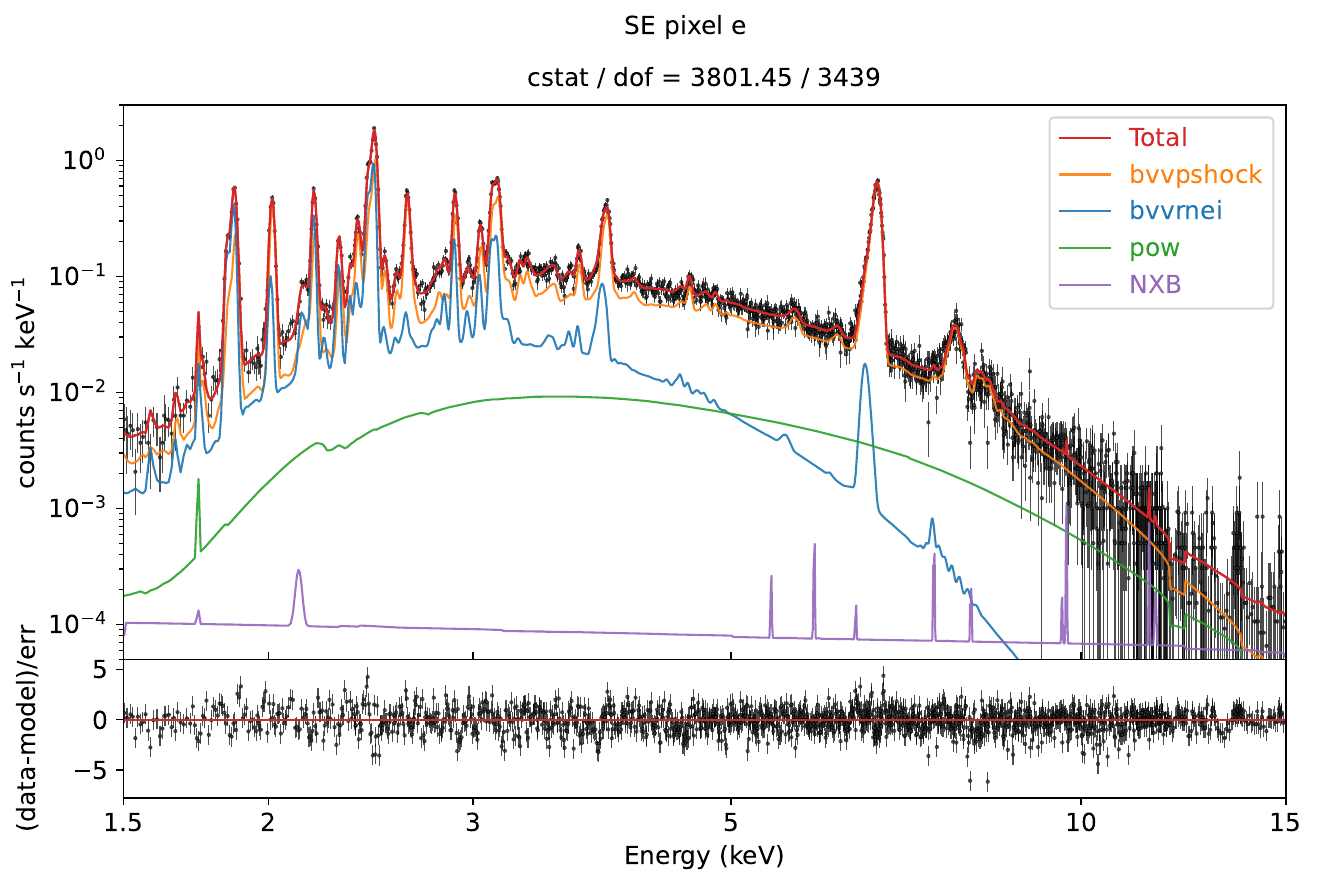} 
\end{center}
\caption{LEFT: Resolve spectra from the on-axis point, so-called super-pixel ``e'' for the SE pointing in the 1.5-15.0 keV range fit with {\texttt{SPEX}}~3.08.00. RIGHT: Resolve spectra from the on-axis point, super-pixel ``e'' for the SE pointing in the 1.5-15.0 keV range fit with {\texttt{SPEX}}~3.08.01$^*$.  The data are represented by the black data points. 
The upper red curve is the total model, the orange curve is the high~kT bvvpshock component, the blue curve is the low~kT bvvrnei component (the {\texttt{SPEX}} equivalent model is called \texttt{neij}), the green curve is the power-law component, the brown line at $\sim3.2$~keV is the additional Gaussian component, and the purple curve is the NXB component. 
The residuals are plotted in the lower panel.
{Alt text: Two line graphs showing the same spectrum fit with different models, {\texttt{SPEX}} 3.08.00 on the left and 3.08.01$^*$ on the right.  Additional lines show the model components separately. The lower panel shows the residuals for the fit. The plots use the same horizontal axis that shows the energy in keV.
The spectral plot has a vertical axis that shows the data and model in units of  ${\mathrm{count~s^{-1}~keV^{-1}}}$. The residual plot shows the residuals in units of data minus the model divided by the error.}
}
\label{fig:se_spex3.08.00_3.08.01}
\end{figure*}

We extracted spectra from the on-axis, $2\times2$ pixel, region labeled as super-pixel ``e'' in Figure~\ref{fig:resolve_pixels} from both the SE and NW observations. These $2\times2$ pixel regions have a relatively high purity of counts arising from these parts of Cas~A as described in section ~\ref{sec:ssm} and are not affected by uncertainties in the off-axis mirror response. The on-axis mirror response should have the smallest uncertainties at this point in the mission. Although it is worth noting that a significant amount of the flux in these regions arises from the adjacent regions as the 
purity values are $\approx48\%$ and $\approx31\%$ for the SE and NW spectra respectively.  Nevertheless, these spectra should exhibit significantly less broadening due to the mixing of emission from different regions as described in the previous section \ref{sec:integrated_spectra} and as shown in Figures~3 \& 8 in \cite{vink2025}.  The SE spectrum is shown in Figures~\ref{fig:se_atomdb3.0.9_3.1.0} and ~\ref{fig:se_spex3.08.00_3.08.01} and the NW spectrum is shown in Figures~\ref{fig:nw_atomdb3.0.9_3.1.0} and~\ref{fig:nw_spex3.08.00_3.08.01} in the 1.5-15.0~keV band.  The spectral models are described in detail in the next section~\ref{sec:spectral_models}.  The SE and NW pixel ``e'' spectra are similar in that the lines of Si, S, Ar, Ca, Fe, \& Ni dominate the flux but the relative contribution of the lines compared to the continua components is different between the two spectra. The line emission is relatively stronger in the SE spectrum than in the NW spectrum. The Si and S lines appear to be narrower than in the spectrum from the entire array shown in Figure~\ref{fig:int_spectrum} but the 
Fe~K emission appears to be relatively broad.
The features around the Ni~{\footnotesize{XXVII}}~He$\alpha$ triplet and the Fe~{\footnotesize{XXV}}~He$\beta$ lines at $\sim7.8$~keV appear weaker relative to the Fe~{\footnotesize{XXV}}~He$\alpha$ triplet in the NW spectrum compared to the SE spectrum. These similarities and differences will be discussed in the next section~\ref{sec:fit_results}.

\subsubsection{Spectral Models}
\label{sec:spectral_models}

Initial fits with a single thermal, non-equilibrium ionization (NEI) model showed that it was not possible to fit the line emission from Si to Fe with such a model. We used the {\texttt{vvrnei}} model in \texttt{XSPEC}~\citep{arnaud1996} which assumes that the plasma starts in collisional ionization equilibrium at an initial temperature specified by the user and a single ionization timescale.  Depending on the value of the initial temperature that is specified, this model can represent a recombining or ionizing plasma.
Fits with two ({\texttt{vvrnei}}) components improved the fit but could not represent the broadening of the lines.  Fits with  two broadended NEI models ({\texttt{bvvrnei}}) improved the fit significantly but there were still large residuals around the Fe~{\footnotesize{XXV}}~He$\alpha$ triplet given its asymmetric profile. We then substituted a {\texttt{bvvphock}} model for one of the {\texttt{bvvrnei} components.  A {\texttt{bvvphock}} model is a constant temperature, plane-parallel shock model that takes into account the distribution of ionization timescales.
The addition of a {\texttt{bvvphock} model provided a significantly improved fit to the Fe~{\footnotesize{XXV}}~He$\alpha$ triplet because it integrates the emission over a range of ionization timescales at a single temperature.  This can produce emission at lower charge states than Fe~{\footnotesize{XXV}} depending on the temperature and ionization timescale and can fill in the energy range just below the peak of the He$\alpha$ triplet.  Based on these results, we selected a two component model consisting of a {\texttt{bvvrnei}} component to provide most of the emission at low energies and a {\texttt{bvvphock}} component to provide most of emission at high energies, with both components contributing significantly in the middle of the band. 
We allowed the abundances of Mg, Si, P, S, Cl, Ar, K, Ca, Ti, Cr, Mn, Fe, \& Ni to vary.  Abundances of elements with lower atomic number than Mg and abundances of Al, V, Co, Cu, \& Zn were frozen at the Solar value.
We adopt the solar abundance values of \cite{lodders2009} for all of the fits in this paper.  For the {\texttt{XSPEC} fits we set the parameter {\texttt{APECBROADPSEUDO}} such that the low-flux lines were also broadened.  A similar effect was accomplished in the {\texttt{SPEX}} fits by broadening all the lines in the model.  
The non-thermal emission was represented by a single power-law component.  The NXB was modeled using the template developed by the \xrism\/ calibration team and normalized to the flux in the 13.0-15.0~keV band. The NXB template consists of a powerlaw with an index of 0.16 and seventeen Gaussians for the instrumental lines. The absorption due to the intervening interstellar medium (ISM) was modeled using the {\texttt{phabs}} component.  We also included a Gaussian component because fits with the pre-launch versions of {\texttt{AtomDB}} and {\texttt{SPEX}} did not include sufficient emission around an energy of $\sim3.2$~keV to account for a line-like feature.   The final model we used to fit the SE and NW spectra consisted of: an absorption component ({\texttt{phabs}}), two thermal components (a  {\texttt{bvvrnei}} and a {\texttt{bvvphock}}), one nonthermal component ({\texttt{powerlaw}}), a Gaussian, and the NXB component.

Detailed fits with this model provided reasonable fits to both data sets across the relatively broad band of 1.5-15.0~keV but revealed significant degeneracies among some of the parameters.  We started the fits with most parameters free to vary but it became clear that we would need to link and/or constrain some of the parameters otherwise the fits would not converge.  We fixed the neutral hydrogen column density (${\mathrm{N_H}}$) to $\mathrm{1.5\times10^{22}~cm^{-2}}$ since the fits are not sensitive to the absorbing column with the Gate Valve closed.  
We started the fits with the Mg, Si, P, S, Cl, Ar, K, Ca, Ti, Cr, Mn, Fe, \& Ni abundances free for each of the two thermal components.  This proved to be too many degrees of freedom for the fits and we linked the abundances for the two thermal components.  The initial temperature of the {\texttt{bvvrnei}} component was set to 0.080~keV since we expect the plasma to be ionizing in Cas~A and the lower limit on the ionization timescale for the {\texttt{bvvpshock}} component was set to 0. The temperature, ionization timescale, normalization, redshift and broadening of each thermal component were allowed to vary separately.  The power-law index was set to 2.8 for the SE spectrum and to 3.4 for the NW spectrum based on the results presented in \cite{helder2008}. The normalization of the powerlaw component was allowed to vary.  The energy and normalization of the Gaussian component were allowed to vary while the width was frozen at 0.02 keV based on the results of fits in a narrow energy range. Finally, the constant factor for the NXB component was determined by fitting in the 13.0-15.0~keV band and then held fixed to that value when the entire model was fit in the 1.5-15.0~keV range. Fits conducted in this manner yielded reasonable results, however, the calculation of the uncertainties on the parameters revealed a degeneracy between some of the parameters for the two thermal components.  Setting the ionization timescale for the {\texttt{bvvrnei}} component to ${\mathrm{1.0\times10^{11}~cm^{-3}~s}}$ removed this degeneracy and allowed the uncertainties to be calculated for the remaining free parameters in the thermal models.  This resulted in 25 free parameters for the fits.

  \begin{longtable}{lcccc}
  \caption{Spectral fit results from the SE Pointing on-axis $2\times2$ pixel region with AtomDB and SPEX }
  \label{tab:spec_se}  
\hline\noalign{\vskip3pt} 
  Parameter & AtomDB 3.0.9 & AtomDB 3.1.0 & SPEX v3.08.00 & SPEX v3.08.01$^*$ \\   [2pt] 
\hline\noalign{\vskip3pt} 
\endfirsthead      
\hline\noalign{\vskip3pt}
Parameter & AtomDB 3.0.9 & AtomDB 3.1.0 & SPEX v3.08.00 & SPEX v3.08.01$^*$ \\   [2pt] 
\hline\noalign{\vskip3pt} 
\endhead
\hline\noalign{\vskip3pt} 
\endfoot
\hline\noalign{\vskip3pt} 
\endlastfoot 
  bvvrnei &  &  &  &  \\
  \hline
  ${\mathrm{N_H}}$ ($\mathrm{\times10^{22}~cm^{-2}}$) & 1.50 & 1.50 & 1.50 & 1.50 \\
  ${\mathrm{n_et~(\times10^{11}~cm^{-3}~s)}}$  & 1.00 & 1.00 & 1.00  & 1.00 \\
  $\mathrm {kT_{init}(keV)}$ & 0.08 & 0.08 & 0.002 & 0.002 \\
  kT(keV) & $1.04^{+0.01}_{-0.01}$ & $0.91^{+0.03}_{-0.01}$ & $0.98^{+0.01}_{-0.01}$  & $0.96^{+0.01}_{-0.01}$ \\
  Mg & $3.71^{+0.62}_{-0.42}$ & $2.46^{+0.43}_{-0.41}$ & $4.47^{+0.41}_{-0.48}$ & $4.71^{+0.46}_{-0.46}$ \\
  Si & $8.88^{+0.50}_{-0.61}$ & $7.46^{+0.45}_{-0.31}$ & $8.21^{+0.09}_{-0.08}$  & $8.63^{+0.09}_{-0.10}$ \\
  P & $13.75^{+1.69}_{-1.63}$ & $10.00^{+1.40}_{-1.44}$ & $11.70^{+2.58}_{-1.08}$ & $11.82^{+1.59}_{-1.55}$\\
  S & $9.64^{+0.49}_{-0.31}$ &  $8.07^{+0.76}_{-0.34}$ & $8.47^{+0.08}_{-0.05}$ & $8.88^{+0.18}_{-0.07}$\\
  Cl & $10.92^{+1.35}_{-1.35}$ & $10.11^{+1.17}_{-1.15}$ & $10.98^{+3.35}_{-0.59}$ & $10.90^{+1.25}_{-1.22}$ \\
  Ar & $8.52^{+0.22}_{-0.31}$  & $7.13^{+0.49}_{-0.30}$ & $7.27^{+0.13}_{-0.11}$ & $7.63^{+0.86}_{-0.12}$ \\
  K & $11.59^{+1.61}_{-1.59}$ & $9.41^{+1.34}_{-1.31}$ & $11.21^{+7.37}_{-0.48}$ & $10.66^{+1.50}_{-1.46}$ \\
  Ca & $9.87^{+1.06}_{-0.81}$  & $7.87^{+0.57}_{-0.34}$ & $8.19^{+0.17}_{-0.15}$ & $8.60^{+0.16}_{-0.16}$ \\
  Ti & $12.63^{+3.85}_{-3.77}$  & $9.93^{+2.99}_{-2.89}$  & $8.95^{+2.68}_{-3.72}$ & $8.99^{+3.39}_{-3.04}$ \\
  Cr & $7.28^{+1.20}_{-1.16}$ & $5.45^{+0.88}_{-0.83}$ & $6.41^{+0.77}_{-1.13}$  & $6.42^{+0.97}_{-0.95}$  \\
  Mn & $8.49^{+2.20}_{-2.20}$ & $5.55^{+1.44}_{-1.37}$ & $7.52^{+1.38}_{-2.37}$  &  $7.48^{+1.93}_{-1.89}$ \\
  Fe & $14.52^{+2.78}_{-1.38}$ & $11.01^{+1.60}_{-0.60}$ & $10.11^{+0.87}_{-0.65}$ & $10.28^{+0.55}_{-0.21}$ \\
  Ni & $30.73^{+1.85}_{-2.36}$ & $11.65^{+1.34}_{-1.26}$ & $9.70^{+0.93}_{-0.80}$ & $6.62^{+0.90}_{-0.89}$ \\
  Redshift (z)($\times10^{-3}$) & $-1.68^{+0.11}_{-0.11}$ & $-1.40^{+0.10}_{-0.12}$ & $-1.15^{+0.01}_{-0.02}$ & $-1.17^{+0.01}_{-0.01}$\\
  $\mathrm{\sigma_v~(km~s^{-1})}$ & $1118^{+21}_{-22}$ & $1074^{+25}_{-24}$ & $1085^{+24}_{-18}$ & $1089^{+21}_{-24}$\\
  AtomDB Norm ($\times10^{-2}~\mathrm{cm^{-5}}$) & $3.48^{+0.17}_{-0.48}$ & $4.23^{+0.21}_{-0.32}$ & $4.33^{+0.06}_{-0.04}$ & $4.12^{+0.09}_{-0.09}$\\
  SPEX Norm ($\times10^{63}~\mathrm{m^{-3}}$) & - & - & $6.00^{+0.06}_{-0.05}$  & $5.70^{+0.12}_{-0.12}$ \\
  \hline
  bvvpshock &  &  &  &  \\
  \hline
  ${\mathrm{n_et~(\times10^{11}~cm^{-3}~s)}}$  & $2.16^{+0.07}_{-0.04}$ & $1.88^{+0.04}_{-0.06}$ & $2.12^{+0.13}_{-0.06}$  &  $2.11^{+0.02}_{-0.03}$ \\
  kT(keV) & $2.84^{+0.06}_{-0.10}$ & $2.79^{+0.06}_{-0.07}$ & $2.79^{+0.02}_{-0.03}$ & $2.73^{+0.07}_{-0.12}$ \\
  Redshift (z)($\times10^{-3}$) & $-4.48^{+0.07}_{-0.07}$  & $-4.35^{+0.04}_{-0.07}$  & $-4.77^{+0.23}_{-0.01}$  & $-4.67^{+0.02}_{-0.01}$ \\
  $\mathrm{\sigma_v~(km~s^{-1})}$ & $1341^{+16}_{-17}$  & $1322^{+8}_{-16}$ & $1270^{+10}_{-15}$  & $1282^{+12}_{-13}$ \\
  AtomDB Norm ($\times10^{-2}\mathrm{cm^{-5}}$) & $1.54^{+0.14}_{-0.13}$  & $2.16^{+0.09}_{-0.16}$ & $2.09^{+0.01}_{-0.01}$ & $2.07^{+0.02}_{-0.22}$ \\
  SPEX Norm ($\times10^{63}~\mathrm{m^{-3}}$) &  - &  - & $2.89^{+0.02}_{-0.01}$ & $2.86^{+0.02}_{-0.31}$ \\
  \hline
  power-law &  &  & &  \\
  \hline
  index & 2.80 & 2.80 & 2.80 &  2.80 \\
  AtomDB Norm ($\times10^{-3}$) & $6.07^{+1.41}_{-1.05}$ & $4.00^{+1.10}_{-0.44}$ & $3.02^{+0.11}_{-0.19}$ & $3.76^{+0.15}_{-0.25}$ \\
  (photons ${\mathrm {cm^{-2} s^{-1}}}$ at 1.0 keV) & &  &  & \\\
  SPEX Norm ($\times10^{42}$) & - & - & $4.17^{+0.16}_{-0.27}$ & $5.19^{+0.21}_{-0.35}$  \\
  (photons ${\mathrm {s^{-1} keV^{-1}} }$ at 1.0 keV) & &  &  & \\
  \hline
  Gaussian &  &  &  & \\
  \hline
  E(keV)  & $3.22^{+0.04}_{-0.03}$  & -  & $3.22^{+0.01}_{-0.01}$  & - \\
  AtomDB Norm ($\times10^{-5}$ photons ${\mathrm {cm^{-2} s^{-1}}}$) & $2.06^{+0.37}_{-0.34}$ & -  & $2.75^{+0.38}_{-0.35}$   &  -  \\  
  SPEX Norm ($\times10^{40}$ photons ${\mathrm s^{-1}}$) & - & - & $3.80^{+0.52}_{-0.48}$ &  -  \\ 
  \hline
  Fit Statistics &  &  &  & \\
  \hline
  C Statistic  & 3821.5  & 3674.5 & 3793.1  & 3801.4 \\
  Degrees of Freedom & 3350 & 3352  & 3437  & 3439  \\ 
  \end{longtable}

\begin{longtable}{lcccc}
  \caption{Spectral fit results from the NW Pointing on-axis $2\times2$ pixel region with AtomdDB and SPEX 
  }\label{tab:spec_nw}  
\hline\noalign{\vskip3pt} 
  Parameter & AtomDB 3.0.9 & AtomDB 3.1.0 & SPEX v3.08.00 & SPEX v3.08.01$^*$ \\   [2pt] 
\hline\noalign{\vskip3pt} 
\endfirsthead      
\hline\noalign{\vskip3pt}
Parameter & AtomDB 3.0.9 & AtomDB 3.1.0 & SPEX v3.08.00 & SPEX v3.08.01$^*$ \\   [2pt] 
\hline\noalign{\vskip3pt} 
\endhead
\hline\noalign{\vskip3pt} 
\endfoot
\hline\noalign{\vskip3pt} 
\endlastfoot 
  bvvrnei &  &  &  &  \\
  \hline
  ${\mathrm{N_H}}$ ($\mathrm{\times10^{22}~cm^{-2}}$) & 1.50 & 1.50 & 1.50 & 1.50 \\
  ${\mathrm{n_et~(\times10^{11}~cm^{-3}~s)}}$  & 1.00 & 1.00 & 1.00  & 1.00 \\
  $\mathrm {kT_{init}(keV)}$ & 0.08 & 0.08 & 0.002 & 0.002 \\
  kT(keV) & $1.08^{+0.05}_{-0.06}$ & $1.05^{+0.06}_{-0.09}$ & $0.99^{+0.34}_{-0.02}$  & $1.10^{+0.06}_{-0.14}$ \\
  Mg & $0.76^{+0.51}_{-0.48}$ & $1.07^{+0.39}_{-0.38}$ & $3.00^{+0.71}_{-0.79}$ & $1.96^{+0.50}_{-0.54}$ \\
  Si & $2.77^{+0.23}_{-0.25}$ & $2.95^{+0.27}_{-0.34}$ & $5.58^{+0.11}_{-0.09}$  & $3.84^{+0.58}_{-0.29}$ \\
  P & $2.01^{+1.29}_{-1.25}$ & $1.33^{+1.42}_{-1.33}$ & $2.03^{+3.83}_{-1.86}$ & $0.96^{+1.70}_{-0.96}$\\
  S & $3.36^{+0.41}_{-0.31}$ &  $3.52^{+0.34}_{-0.40}$ & $6.25^{+0.07}_{-0.07}$ & $4.26^{+0.54}_{-0.62}$\\
  Cl & $3.25^{+1.05}_{-1.03}$ & $3.84^{+1.15}_{-1.14}$ & $5.30^{+2.20}_{-2.33}$ & $3.67^{+1.29}_{-1.27}$ \\
  Ar & $3.15^{+0.37}_{-0.31}$  & $3.25^{+0.37}_{-0.37}$ & $5.43^{+0.14}_{-0.13}$ & $3.78^{+0.58}_{-0.29}$ \\
  K & $3.54^{+1.21}_{-1.20}$ & $3.75^{+1.24}_{-1.22}$ & $8.75^{+2.93}_{-2.88}$ & $4.30^{+1.48}_{-1.44}$ \\
  Ca & $3.99^{+0.46}_{-0.37}$  & $3.95^{+0.41}_{-0.44}$ & $6.51^{+0.18}_{-0.18}$ & $4.60^{+0.26}_{-0.13}$ \\
  Ti & $1.02^{+2.87}_{-1.02}$  & $0.88^{+2.84}_{-0.88}$  & $0.46^{+4.65}_{-0.46}$ & $0.64^{+2.52}_{-0.64}$ \\
  Cr & $3.07^{+0.90}_{-0.88}$ & $3.19^{+0.85}_{-0.83}$ & $4.96^{+1.27}_{-1.29}$  & $3.47^{+0.94}_{-0.94}$  \\
  Mn & $0.74^{+1.67}_{-0.74}$ & $0.54^{+1.39}_{-0.54}$ & $2.50^{+2.89}_{-2.50}$  &  $1.34^{+2.12}_{-1.34}$ \\
  Fe & $3.90^{+0.28}_{-0.29}$ & $3.85^{+0.36}_{-0.28}$ & $5.21^{+0.06}_{-0.06}$ & $3.80^{+0.49}_{-0.37}$ \\
  Ni & $3.67^{+0.82}_{-0.85}$ & $1.04^{+0.77}_{-0.74}$ & $1.50^{+0.78}_{-0.79}$ & $0.53^{+0.59}_{-0.53}$ \\
  Redshift (z)($\times10^{-3}$) & $7.80^{+0.20}_{-0.36}$ & $7.59^{+0.47}_{-0.46}$ & $11.04^{+3.01}_{-0.08}$ & $8.90^{+0.28}_{-0.69}$\\
  $\mathrm{\sigma_v~(km~s^{-1})}$ & $2075^{+78}_{-77}$ & $2067^{+88}_{-88}$ & $1050^{+55}_{-115}$ & $2025^{+70}_{-169}$\\
  AtomDB Norm ($\times10^{-2}~\mathrm{cm^{-5}}$) & $5.72^{+0.90}_{-0.76}$ & $4.87^{+0.91}_{-0.57}$ & $1.91^{+0.04}_{-0.07}$ & $3.89^{+0.75}_{-0.57}$ \\
  SPEX Norm ($\times10^{63}~\mathrm{m^{-3}}$) &  &  & $2.64^{+0.06}_{-0.10}$  & $5.38^{+1.04}_{-0.79}$ \\
  \hline
  bvvpshock &  &  &  &  \\
  \hline
  ${\mathrm{n_et~(\times10^{11}~cm^{-3}~s)}}$  & $1.17^{+0.01}_{-0.03}$ & $1.07^{+0.02}_{-0.03}$ & $1.05^{+0.01}_{-0.01}$  &  $1.10^{+0.01}_{-0.02}$ \\
  kT(keV) & $3.87^{+0.07}_{-0.10}$ & $3.77^{+0.11}_{-0.12}$ & $3.62^{+0.01}_{-0.02}$ & $3.73^{+0.19}_{-0.10}$ \\
  Redshift (z)($\times10^{-3}$) & $6.38^{+0.22}_{-0.20}$  & $6.47^{+0.21}_{-0.20}$  & $5.36^{+0.90}_{-1.21}$  & $5.86^{+0.06}_{-0.08}$ \\
  $\mathrm{\sigma_v~(km~s^{-1})}$ & $1868^{+39}_{-39}$  & $1856^{+39}_{-35}$ & $1830^{+24}_{-25}$  & $1854^{+53}_{-26}$ \\
  AtomDB Norm ($\times10^{-2}\mathrm{cm^{-5}}$) & $3.25^{+0.14}_{-0.29}$  & $3.31^{+0.31}_{-0.22}$ & $2.32^{+0.10}_{-0.21}$ & $2.94^{+0.17}_{-0.32}$  \\
  SPEX Norm ($\times10^{63}~\mathrm{m^{-3}}$) & -  & - & $3.21^{+0.02}_{-0.02}$ & $4.06^{+0.24}_{-0.44}$  \\
  \hline
  power-law &  &  & &  \\
  \hline
  index & 3.40 & 3.40 & 3.40 &  3.40 \\
  AtomDB Norm ($\times10^{-2}$) & $6.89^{+0.76}_{-0.59}$ & $7.09^{+0.65}_{-0.88}$ & $9.98^{+0.50}_{-0.65}$ & $8.03^{+0.94}_{-0.65}$ \\
  (photons ${\mathrm {cm^{-2} s^{-1}}}$ at 1.0 keV) & &  &  & \\\
  SPEX Norm ($\times10^{44}$) & - & - & $1.38^{+0.07}_{-0.09}$  & $1.11^{+0.13}_{-0.09}$  \\
  (photons ${\mathrm {s^{-1} keV^{-1}} }$ at 1.0 keV) & &  &  & \\
  \hline  
  Gaussian &  &  &  & \\
  \hline
  E(keV)  & $3.19^{+0.01}_{-0.01}$  & -  & $3.19^{+0.01}_{-0.01}$  & - \\
  AtomDB Norm ($\times10^{-5}$ photons ${\mathrm {cm^{-2} s^{-1}}}$) & $1.34^{+0.54}_{-0.53}$ & -  &  $1.69^{+0.54}_{-0.54}$ &  - \\  
  SPEX Norm ($\times10^{40}$ photons ${\mathrm s^{-1}}$) & - & - & $2.34^{+0.75}_{-0.75}$  &  -  \\ 
  \hline
  Fit Statistics &  &  &  & \\
  \hline
  C Statistic  & 4001.9  & 4004.9 & 4041.5  & 4049.2 \\
  Degrees of Freedom & 3423 & 3425  & 3533  & 3535  \\  
  \end{longtable}

\section{Discussion}
\label{sec:discuss}

Fits were performed with the versions of {\texttt{AtomDB}} (3.0.9) and {\texttt{SPEX}} (3.08.00) that were available before the launch of \xrism\, to provide a baseline of the state of our knowledge before the \xrism\, data became available.  We then performed fits with the versions of {\texttt{AtomDB}} (3.1.0) and {\texttt{SPEX}} (3.08.01$^*$) that were released after launch partially in response to some of the issues raised by the high-resolution data from Resolve. We used {\texttt{SPEX}} version 3.08.01 with SPEXACT version 3.08.02, which is the atomic database to be published with the updated {\texttt{SPEX}} software (3.08.02), and hereafter refer to it as {\texttt{SPEX}} 3.08.01$^*$.  Fits with both versions of {\texttt{AtomDB}} were performed with {\texttt{XSPEC}} version~12.14.1.

{\texttt{SPEX}} version 3.08.01$^*$ introduces significant updates to the atomic database compared to version 3.08.00, including:

\begin{enumerate}
    \item \textbf{Rydberg Series}: Calculation of hydrogen-like and helium-like S and Fe ions in the Rydberg series up to \( n = 52 \).
    \item \textbf{Updated Li-, Be-, and B-like ions}: More comprehensive innershell and dielectronic recombination (DR) calculations for ions in Li-, Be-, and B- like isoelectronic sequences.
    \item \textbf{Expanded Innershell Lines for Low Charge States}: A more complete dataset of innershell lines for Cr, Mn, Fe, and Ni up to P-like configurations.
     \item \textbf{Doppler Broadening}: Incorporation of Doppler broadening for the recombination continuum and absorption edges.
\end{enumerate}

AtomDB version 3.1.0 (and subsequent minor updates to 3.1.3, as of the time of writing) included several changes to the underlying atomic data, both improvements and bug fixes.

\begin{enumerate}
    \item \textbf{Rydberg Series:} Calculation of H- and He-like emission for the Rydberg series transitions ($n\rightarrow1$) for all elements, up to $n=50$. Previously this data stopped at $n=10$.
    \item \textbf{Fix to K$\beta$ from Li-like ions:} Inner shell emission from Li-like ions was taken from the literature but not correctly handled. This led to $K\beta$ being completely suppressed in these ions. This has now been corrected for all Li-like ions.
    \item \textbf{DR satellite lines} Previously, the satellites of Fe were only in the model for $1s 2l nl' \rightarrow 1s^2 2l$  transitions where $n\le 5$. These have been augmented by increasing $n$ to 15, and including $1s [3,4]l nl' \rightarrow 1s^2 [3,4]l$.
    \item \textbf{Ionization rates} The data for ionization was updated from \citet{Bryans2009} to \citet{Urdampilleta2017a}. In addition, multiple ionization was added from \citet{Hahn2017Electron-impactZinc}

\end{enumerate}

We compare the results from the different versions of {\texttt{AtomDB}} and {\texttt{SPEX}} in the next section~\ref{sec:fit_results}.

\subsection{Spectral Fit Results}
\label{sec:fit_results}

The fit results for the SE super-pixel ``e'' spectrum with {\texttt{AtomDB}}~3.0.9 and~3.1.0 and {\texttt{SPEX}}~3.08.00 and~3.08.01$^*$ for the model described above are listed in Table~\ref{tab:spec_se}.   The spectra and model for the fits with {\texttt{XSPEC}} and {\texttt{AtomDB}} are shown in Figure~\ref{fig:se_atomdb3.0.9_3.1.0} and the fits for {\texttt{SPEX}} are shown in Figure~\ref{fig:se_spex3.08.00_3.08.01} . We first compare fit results with {\texttt{AtomDB}}~3.0.9 to those with {\texttt{AtomDB}}~3.1.0, then the results with {\texttt{SPEX}}~3.08.00 to {\texttt{SPEX}}~3.08.01$^*$, and finally the results with {\texttt{AtomDB}}~3.1.0 to {\texttt{SPEX}}~3.08.01$^*$. 

The fits with {\texttt{AtomDB}}~3.0.9 and 3.1.0 adequately represent the data across the entire bandpass as can be seen by the residuals in Figure~\ref{fig:se_atomdb3.0.9_3.1.0}, but some of the bright lines have significant residuals.
The value of the C-statistic decreases from 3821.5 with 3350 degrees of freedom (DOF) to 3674.5  (3352 DOF) for the fit with {\texttt{AtomDB}}~3.1.0 compared to 3.0.9, indicating a significantly improved fit with 3.1.0. The DOF are lower for the fit with {\texttt{AtomDB}}~3.1.0 since the Gaussian at 3.2~keV is not included in that model. Notably all of the abundances are enhanced significantly compared to Solar values. This is consistent with a strong ejecta component contribution to this region of the remnant.  The Ar~{\footnotesize{XVII}} He$\alpha$ ($\sim3.1$~keV) and Ca~{\footnotesize{XIX}} He$\alpha$ ($\sim3.9$~keV) line complexes are also clearly visible in these spectra in addition to the lines from Si, S, Fe, \& Ni.
The two thermal components both contribute significantly to the Si and S line emission from 1.5-3.0~keV, but the contribution of the {\texttt{bvvrnei}} component diminishes with increasing energy such that the {\texttt{bvvpshock}} component provides most of the emission for the Fe~K and Ni~K complexes.  The power-law component does not become comparable to the thermal continuum until an energy of $\sim11$~keV.
There is overall good agreement between the fitted parameters derived with {\texttt{AtomDB}}~3.0.9 and 3.1.0, with some notable exceptions. All of the abundances agree to within $2.0\sigma$ with some agreeing to within $1.0\sigma$, except for the Ni abundance which differs at the many $\sigma$ level.
The abundances are systematically lower with {\texttt{AtomDB}}~3.1.0 than with 3.0.9. This is primarily driven by the increase in the normalizations for the two thermal components relative to the power-law component. The normalizations for the {\texttt{bvvrnei}} and {\texttt{bvvpshock}} components are $22\%$ and $40\%$ higher with {\texttt{AtomDB}}~3.1.0, while the normalization for the power-law component is $35\%$ lower.  The increase in the normalizations for the thermal components results in more of the continuum being attributed to the thermal components, which reduces the contrast between the line emission and the continuum emission resulting in lower abundances to explain the line emission. In addition, the temperature and redshift for the  {\texttt{bvvrnei}} component and the ionization timescale for the {\texttt{bvvpshock}} component are significantly lower with {\texttt{AtomDB}}~3.1.0.  All other parameters agree at the $1.0\sigma$ level, notably the broadening terms for both thermal components and the temperature and redshift for the \texttt{bvvpshock}} component.
Of all the parameters, the Ni abundance shows the largest difference between {\texttt{AtomDB}}~3.0.9 and 3.1.0, decreasing from $30.73^{+1.85}_{-2.36}$ to $11.65^{+1.34}_{-1.26}$.  This difference is discussed in more detail in section~\ref{sec:Ni_abundnance}.

The fits with {\texttt{SPEX}}~3.08.00 and 3.08.01$^*$ likewise represent the data well across the entire bandpass as can be seen by the residuals in Figure~\ref{fig:se_spex3.08.00_3.08.01}.
The value of the C-statistic increases from 3793.1 (3437 DOF) to 3801.4  (3439 DOF) for the fit with {\texttt{SPEX}}~3.08.01$^*$ compared to 3.08.00, indicating a slightly worse fit with 3.08.01$^*$.
There is overall excellent agreement for the fitted parameters derived from the fits with the two versions of {\texttt{SPEX}}.  All of the best-fitted parameters agree within $2.0\sigma$ except for the abundances of Si \& S, and the redshift for the {\texttt{SPEX}} equivalent of the  {\texttt{bvvpshock}} component.  The Si \& S abundances agree at the $3.0\sigma$ level but the redshift for the {\texttt{bvvpshock}} component disagrees at more than the $3.0\sigma$ level. The Ni abundance is lower with {\texttt{SPEX}}~3.08.01$^*$ but agrees at the $2.0\sigma$ level.  The abundances with {\texttt{SPEX}}~3.08.01$^*$ are systematically higher than with 3.08.00.  In contrast to the fits with {\texttt{AtomDB}}, the normalizations for the thermal components are lower with {\texttt{SPEX}}~3.08.01$^*$ and the normalization for the power-law component is higher, resulting in higher abundances for the thermal components. 
The uncertainties on the redshifts in the {\texttt{SPEX}} models can be small and sometimes asymmetric. We explored the redshift parameter space for both components in detail and found it to be complex with many local minima.  This exploration confirmed the small values of the uncertainties in some cases and the asymmetric nature of the uncertainties in others. The {\texttt{bvvpshock}} redshift only differs by $2\%$ between the \texttt{SPEX}}~3.08.00 and 3.08.01$^*$ fits, but given the small uncertainties this difference is significant.

The differences in the fitted parameters between {\texttt{AtomDB}}~3.1.0 and {\texttt{SPEX}}~3.08.01$^*$ are perhaps more interesting since they reflect the current state-of-the-art in the knowledge of astrophysical plasma emission models. It should be noted that the number of the degrees of freedom are different between the {\texttt{AtomDB}} and {\texttt{SPEX}} fits due to differences in how the optimal binning algorithm is implemented in {\texttt{ftgrouppha}} and {\texttt{SPEX}}. In general there is good agreement between {\texttt{AtomDB}} and {\texttt{SPEX}} for most of the parameters, with a few exceptions. 
All the abundances agree at the $1.0\sigma$ level, except for Mg, Si, \& Ni which agree at the $3.0\sigma$ level.  As noted previously, the Ni abundance changed significantly between {\texttt{AtomDB}}~3.0.9 and 3.1.0 but even with that change the Ni abundance is still $3.0\sigma$ different between {\texttt{AtomDB}} and {\texttt{SPEX}}.  The Mg abundances disagree at more than the $2.0\sigma$ level, however this measurement is based almost entirely on the Mg~{\footnotesize{XII}} Ly$\beta$ line which is weak in the observed spectrum due to the closed Gate Valve and confused with the neutral Si line from the detector background.  Therefore, the uncertainties on the Mg abundances might be larger than the statistical uncertainties alone given how low the effective area is at this energy with a possibly larger systematic error and significant contribution from the detector background with its associated systematic error.  The normalizations for the {\texttt{bvvrnei}}, {\texttt{bvvpshock}}, and power-law components all agree to within $1.0\sigma$ for {\texttt{AtomDB}}~3.1.0 and {\texttt{SPEX}}~3.08.01$^*$ indicating remarkably good agreement for the overall shape of the spectra.
However, there are differences that affect the details of the spectra.  The redshifts for the {\texttt{bvvrnei}} and  {\texttt{bvvpshock}} components differ at the  many $\sigma$ level, again partially due to the  relative uncertainties being much smaller for the {\texttt{SPEX}} spectral fits.
The temperatures for the {\texttt{bvvrnei}} components agree at the $2.0\sigma$ level but the ionization timescales for the {\texttt{bvvpshock}} components disagree at the many $\sigma$ level.
The broadenings ($\mathrm{\sigma_v}$) for the {\texttt{bvvrnei}} and {\texttt{bvvpshock}} components agree at the $1.0\sigma$ and $2.0\sigma$ levels respectively.
It is interesting to note that the redshift for the {\texttt{bvvrnei}} component in the {\texttt{SPEX}} fit is less negative than the redshift in the {\texttt{AtomDB}} fit but the redshift for the {\texttt{bvvpshock}} component in the {\texttt{SPEX}} fit is more negative than the redshift in the {\texttt{AtomDB}} fit. This would have the effect of broadening the lines more in the {\texttt{SPEX}} fit relative to the {\texttt{AtomDB}} fit.  However, the broadening term for the {\texttt{bvvpshock}} component in the {\texttt{SPEX}} fit is less than that in the {\texttt{AtomDB}} fit.  Therefore, it appears that the redshift and broadening are combining in different ways for the {\texttt{SPEX}} and {\texttt{AtomDB}} fits to represent the widths of the line features.
Overall the agreement between {\texttt{AtomDB}}~3.1.0 and {\texttt{SPEX}}~3.08.01$^*$ is quite good for the SE spectrum except for the redshifts and the Ni abundances.

The fit results for the NW super-pixel ``e'' spectrum with {\texttt{AtomDB}}~3.0.9 and~3.10 and {\texttt{SPEX}}~3.08.00 and~3.08.01$^*$ for the model described above are listed in Table~\ref{tab:spec_nw}.   The spectra and model for the fits with {\texttt{XSPEC}} and {\texttt{AtomDB}} are shown in Figure~\ref{fig:nw_atomdb3.0.9_3.1.0} and the fits with {\texttt{SPEX}} are shown in Figure~\ref{fig:nw_spex3.08.00_3.08.01}.

The fits with {\texttt{AtomDB}}~3.0.9 and 3.1.0 adequately represent the data across the entire bandpass as can be seen by the residuals in Figure~\ref{fig:nw_atomdb3.0.9_3.1.0} and the resulting fit statistic values of C~statistic of $\sim4000$ with 3423/3425 DOF. The value of the C~statistic only changes by 2.5 between the fits indicating that they are of equal quality. Notably the abundances are significantly lower than for the SE spectrum, but they are still enhanced compared to Solar except for Mg, Ti, Mn, and Ni. This would be consistent with an ejecta contribution in this region but not as strong as in the SE spectrum and with a different composition.
All of the fitted parameters (including the model normalizations) agree to within $1.0\sigma$ between {\texttt{AtomDB}}~3.0.9 and 3.1.0 except for the Ni abundance and the ionization timescale for the {\texttt{bvvpshock}} component.  The Ni abundance decreases from $3.67^{+0.82}_{-0.85}$ to $1.04^{+0.77}_{-0.74}$, consistent at the $2.0\sigma$ level.  The ionization timescale for the {\texttt{bvvpshock}} component is $1.17^{+0.01}_{-0.03}{\mathrm{\times10^{11}~cm^{-3}~s}}$ for {\texttt{AtomDB}}~3.0.9 and $1.07^{+0.02}_{-0.03}{\mathrm{\times10^{11}~cm^{-3}~s}}$, differing by $9\%$ and consistent at the $2.0\sigma$ level.
It is possible than there is better agreement between {\texttt{AtomDB}}~3.0.9 and 3.1.0 for the NW spectrum because the line emission is relatively weaker compared to the continuum than in the SE spectrum and therefore the model for the line emission does not contribute as much to the fit statistic.

The fits with {\texttt{SPEX}}~3.08.00 and 3.08.01$^*$ show a larger variation between the fitted parameters as can be seen in Table~\ref{tab:spec_nw}.  The C~statistic is marginally higher, 4049.2 versus 4041.5, for 3.08.01$^*$ compared to 3.08.00. There are several parameters that disagree at more than the $2.0\sigma$ level between these two fits.  Most notable among these are the redshift, broadening, and normalization for the {\texttt{bvvrnei}} component.  The normalization difference can be easily seen in the left and rights panels of Figure~\ref{fig:nw_spex3.08.00_3.08.01} as the blue curve (representing the {\texttt{bvvrnei}} component) is significantly lower ($\sim50\%$) in the fit with {\texttt{SPEX}}~3.08.00.  The normalization of the power-law component is $\sim22\%$ higher for the fit with {\texttt{SPEX}}~3.08.00.  Therefore, the fit with {\texttt{SPEX}}~3.08.00 modeled more of the emission with the power-law component than the {\texttt{bvvrnei}} component.
The broadening with {\texttt{SPEX}}~3.08.00 is approximately half of the value with 3.08.01$^*$ for the low temperature {\texttt{bvvrnei}} component and the redshift is $\sim25\%$ higher with {\texttt{SPEX}}~3.08.00.  This $25\%$ difference is highly significant given the relatively small uncertainties. As noted earlier, the uncertainties on the redshift parameter in the {\texttt{SPEX}} can be small and asymmetric.
The agreement for the parameters of the {\texttt{bvvpshock}} component is better compared to that observed for the {\texttt{bvvrnei}} component, as all parameters agree to within $2.0\sigma$ between {\texttt{SPEX}}~3.08.00 and 3.08.01$^*$.
It appears again that the redshift and the broadening are combining in different ways to produce the observed broadening of the lines.  The {\texttt{bvvrnei}} component has a smaller broadening but a larger redshift for the fit with {\texttt{SPEX}}~3.08.00 compared to the fit with {\texttt{SPEX}}~3.08.01$^*$.  The larger difference between the redshifts for the  {\texttt{bvvrnei}}  and {\texttt{bvvphsock}}  components in the fit with {\texttt{SPEX}}~3.08.00 has the effect of broadening the lines, hence the smaller value of the broadening parameter. 
It is interesting that the fit with {\texttt{SPEX}}~3.08.00 found such a different set of best-fitted parameters than the fit with {\texttt{SPEX}}~3.08.01$^*$. 
However, the parameter space for the fit statistic is complicated with multiple local minima and it is possible that a more advanced algorithm would find a different best fit that would result in better agreement between these two versions of {\texttt{SPEX}}. Overall the differences between the fitted parameters for the NW spectrum with {\texttt{SPEX}}~3.08.00 and 3.08.01$^*$ are larger than for the SE spectrum.

The differences in the best-fitted parameters between {\texttt{AtomDB}}~3.1.0 and \texttt{{SPEX}}~3.08.01$^*$ are smaller for the NW spectrum than for the SE spectrum.
All of the fitted parameters agree at the $1.0\sigma$ level including the normalizations for the {\texttt{bvvrnei}}, {\texttt{bvvpshock}} and power-law components and the Ni abundance, except for the Si \& Ca abundances and the redshift for the {\texttt{bvvpshock}} component. There is excellent agreement for the temperatures and the ionization timescales for the thermal components. The Si \& Ca abundances do agree at the $2.0\sigma$ level, so taken together with the $1.0\sigma$ agreement for the other elements this represents remarkably good agreement between {\texttt{AtomDB}}~3.1.0 and \texttt{{SPEX}}~3.08.01$^*$.  The Ni abundances have relatively large uncertainties such that even though the best-fitted values differ by a factor of two, they still agree within $1.0\sigma$.  The largest difference is in the redshifts for the {\texttt{bvvpshock}} components that disagree at the $2.0\sigma$ level but agree at the $3.0\sigma$ level.  This is somewhat surprising as it should produce a noticeable shift in the line energies, but given the degeneracy between the redshifts and broadenings of the two thermal components the combined effect may not be large.  Even though the normalizations agree at $1.0\sigma$, the differences are notable. The {\texttt{bvvrnei}} and {\texttt{bvvpshock}} normalizations are $25\%$ and $13\%$ higher respectively with {\texttt{AtomDB}} than with SPEX and the power-law normalization is $12\%$ lower with {\texttt{AtomDB}} than with SPEX.  The {\texttt{AtomDB}} fit puts more of the emission in the thermal components and the {\texttt{SPEX}} fit puts more of the emission in the powerlaw component. This has the effect of the abundances being systematically higher in the {\texttt{SPEX}} fits, but the Fe abundances are nearly identical and the Ni abundance is higher in the {\texttt{AtomDB}} fit.  Overall, the agreement between {\texttt{AtomDB}}~3.1.0 and \texttt{{SPEX}}~3.08.01$^*$ for the NW spectrum is excellent.

\begin{figure*}
\begin{center}
\includegraphics[clip, trim =0mm 10mm 50mm 25mm, width=8.5cm,angle=0]
{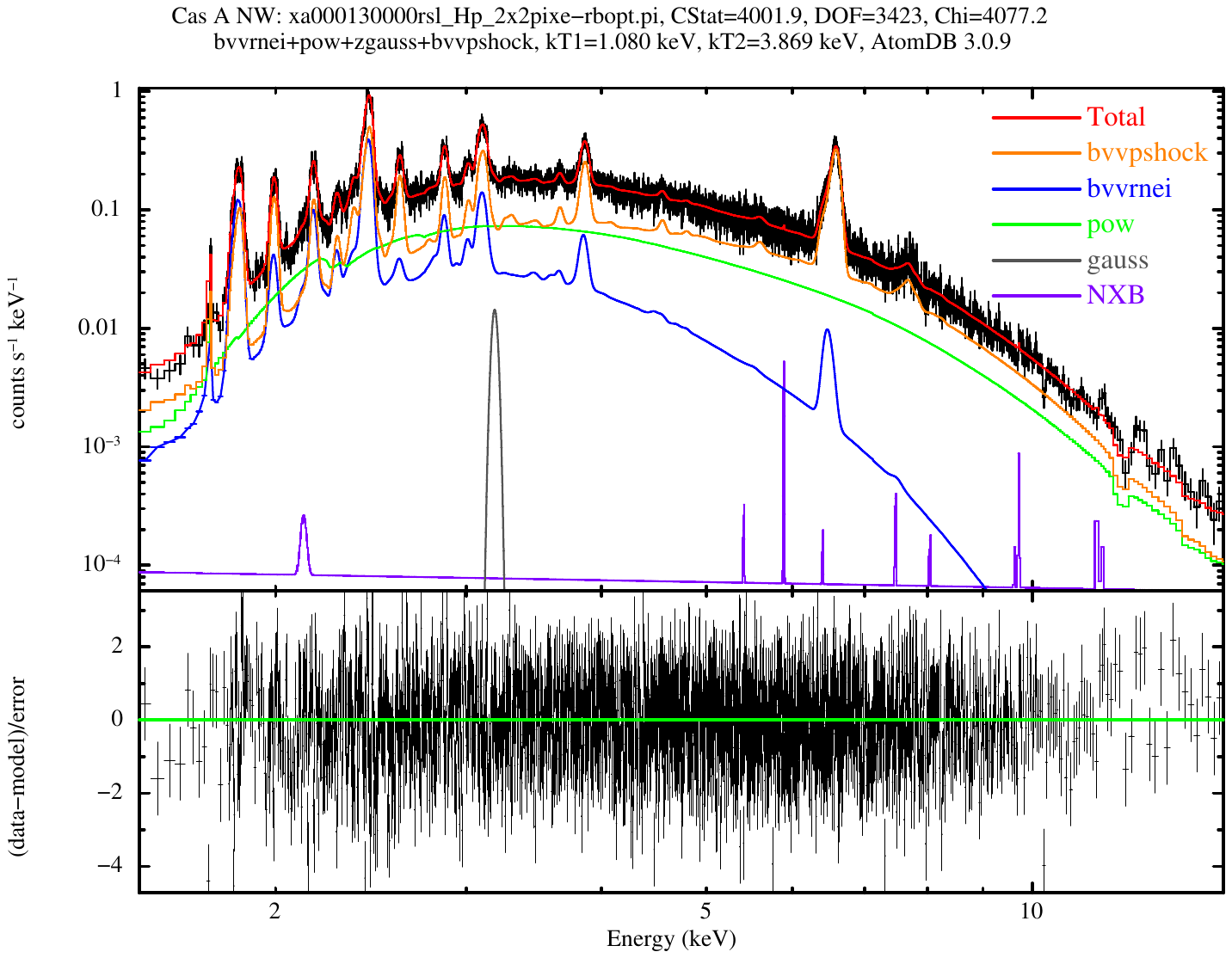} 
\includegraphics[clip, trim =0mm 10mm 50mm 25mm, width=8.5cm,angle=0]{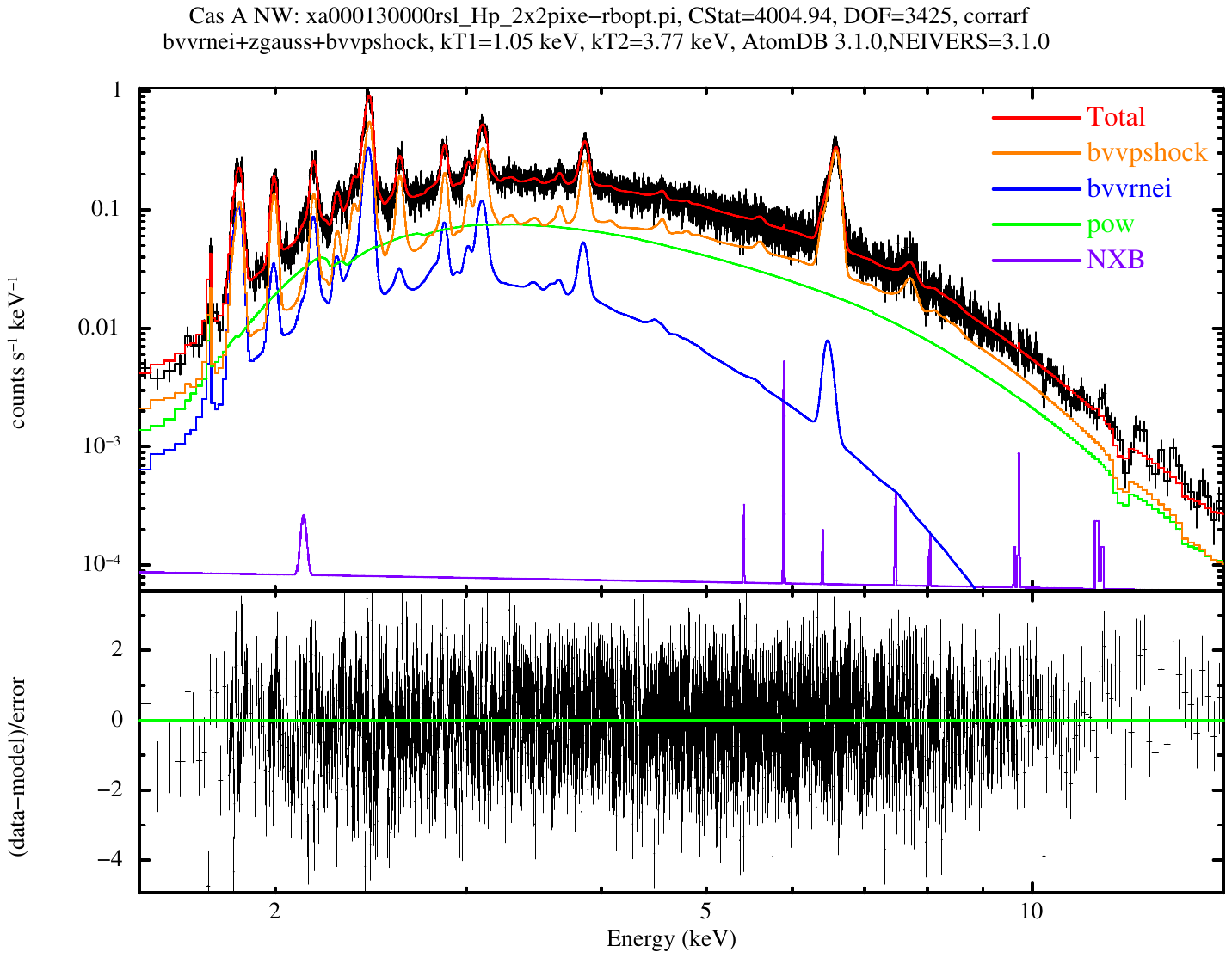} 
\end{center}
\caption{LEFT: Resolve spectra from the on-axis point, so-called super-pixel ``e'' for the NW pointing in the 1.5-15.0 keV range fit with XSPEC and AtomDB~3.0.9. RIGHT: Resolve spectra from the on-axis point, super-pixel ``e'' for the NW pointing in the 1.5-15.0 keV range fit with XSPEC and AtomDB~3.1.0. 
The data are represented by the black data points and have been rebinned for display purposes only. 
The upper red curve is the total model, the orange curve is the high~kT bvvpshock component, the blue curve is the low~kT bvvrnei component, the green curve is the power-law component, the gray line at $\sim3.2$~keV is the additional Gaussian component, and the purple curve is the NXB component. The residuals are plotted in the lower panel.
{Alt text: Two line graphs showing the same spectrum fit with different models, AtomDB 3.0.9 on the left and 3.1.0 on the right.  Additional lines show the model components separately. The lower panel shows the residuals for the fit. The plots use the same horizontal axis that shows the energy in keV.
The spectral plot has a vertical axis that shows the data and model in units of  ${\mathrm{count~s^{-1}~keV^{-1}}}$. The residual plot shows the residuals in units of data minus the model divided by the error.}
}
\label{fig:nw_atomdb3.0.9_3.1.0}
\end{figure*}

\begin{figure*}
\begin{center}
 \includegraphics[clip, trim =0mm 0mm 0mm 15mm, width=8.5cm,angle=0]{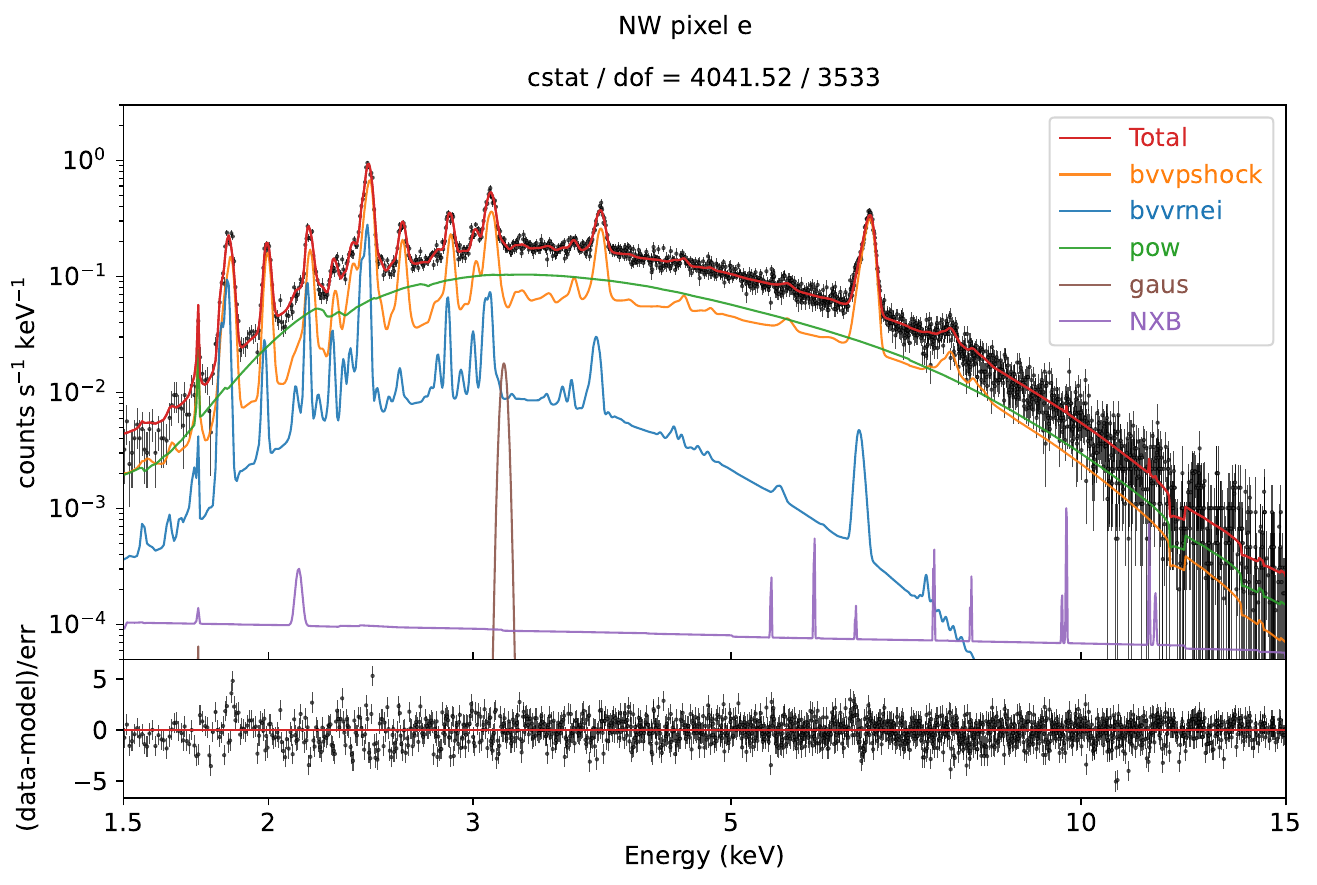} 
\includegraphics[clip, trim = 0mm 0mm 0mm 15mm, width=8.5cm,angle=0]{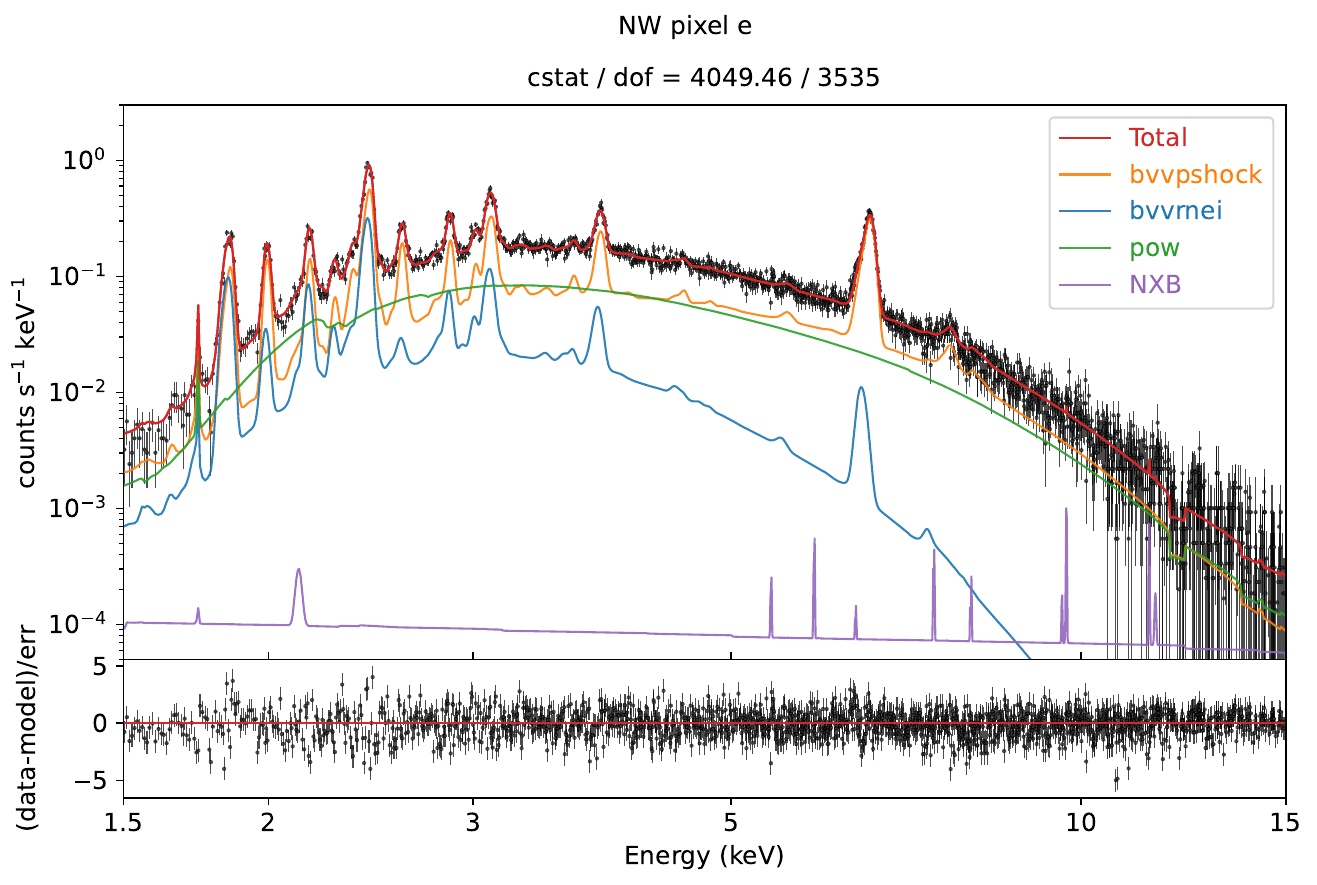} 
\end{center}
\caption{LEFT: Resolve spectra from the on-axis point, so-called super-pixel ``e'' for the NW pointing in the 1.5-15.0 keV range fit with {\texttt{SPEX}}~3.08.00. RIGHT: Resolve spectra from the on-axis point, super-pixel ``e'' for the NW pointing in the 1.5-15.0 keV range fit with {\texttt{SPEX}}~3.08.01$^*$.  The data are represented by the black data points.  
The upper red curve is the total model, the orange curve is the high~kT bvvpshock component, the blue curve is the low~kT bvvrnei component (the {\texttt{SPEX}} equivalent model is called \texttt{neij}), the green curve is the power-law component, the brown line at $\sim3.2$~keV is the additional Gaussian component, and the purple curve is the NXB component.
The residuals are plotted in the lower panel.
{Alt text: Two line graphs showing the same spectrum fit with different models, {\texttt{SPEX}} 3.08.00 on the left and 3.08.1 on the right.  Additional lines show the model components separately. The lower panel shows the residuals for the fit. The plots use the same horizontal axis that shows the energy in keV.
The spectral plot has a vertical axis that shows the data and model in units of  ${\mathrm{count~s^{-1}~keV^{-1}}}$. The residual plot shows the residuals in units of data minus the model divided by the error.}
}
\label{fig:nw_spex3.08.00_3.08.01}
\end{figure*}

\subsection{The Ni Abundance}
\label{sec:Ni_abundnance}

The abundance of Ni (${\mathrm{^{58}Ni}}$ ) is of great interest in the study of SNe and SNRs because the amount of Ni synthesized can be a probe of the environment at the center of the SN explosion \citep{2006ApJ...637..415F,jerkstrand2015,2019MNRAS.488L.114F}.
As noted earlier, the Ni abundance derived from the fits with {\texttt{AtomDB}}~3.0.9 and 3.1.0 changed significantly while the Fe abundance changed by a relatively smaller amount. It decreased from $30.73^{+1.85}_{-2.36}$ to $11.65^{+1.34}_{-1.26}$  in the SE spectral fit and 
from $3.67^{+0.82}_{-0.85}$ to $1.04^{+0.77}_{-0.74}$ in the NW spectral fit.
This change is largely due to the corrections made to K$\beta$ line emission as described in Section~\ref{sec:discuss}. The Fe K$\beta$ lines of the Li-like stage (which overlap the Ni K$\alpha$) had been effectively turned off by an error, and so Ni emission was trying to fill that hole in {\texttt{AtomDB}}~3.0.9. With this corrected in {\texttt{AtomDB}}~3.1.0, less Ni emission is required to compensate for this, lowering the Ni abundance. Other changes (more DR satellites, for example) had a similar but much lesser effect on these data as we are using an ionizing plasma model. The difference is smaller for the results with {\texttt{SPEX}}~3.08.00 and 3.08.01$^*$.  The Ni abundance decreased from $9.70^{+0.93}_{-0.80}$ to $6.62^{+0.90}_{-0.89}$  in the SE spectral fit and from $1.50^{+0.78}_{-0.79}$ to $0.53^{+0.59}_{-0.53}$ in the NW spectral fit.
This is likely due to the fact that the innershell lines and the associated excitation, dielectronic recombination, and innershell ionization rates for low- to mid-charge Ni ions have been added to {\texttt{SPEX}}~3.08.01$^*$. The inclusion of new lines naturally leads to a lower abundance.
 Figure~\ref{fig:fe_ni_atomdb_spex} shows the differences between {\texttt{AtomDB}}~3.1.0 and {\texttt{SPEX}}~3.08.01$^*$ in the energy range containing the Ni~{\footnotesize{XXVII}} He$\alpha$ complex at $\sim7.8$~keV (with a significant contribution from Fe~{\footnotesize{XXV}} He$\beta$ at $\sim7.9$~keV) and the Fe~{\footnotesize{XXVI}} Ly$\beta$ lines at $\sim8.25$~keV for a model spectrum with the parameters of the SE super-pixel ``e'' in the left panel and those models folded through the Resolve response and compared to the data in the right panel. The agreement is quite good for the Ni~{\footnotesize{XXVII}} He$\alpha$ complex and the Fe~{\footnotesize{XXVI}} Ly$\beta$ lines but there are differences at the energies between these main complexes presumably due to the current uncertainties in the strengths of the innershell and dielectronic recombination satellite lines of Fe~{\footnotesize{XXII}},  Fe~{\footnotesize{XXIII}}, and Fe~{\footnotesize{XXIV}}.  This issue requires further investigation and may be addressed in future releases of {\texttt{AtomDB}} and {\texttt{SPEX}} to resolve this apparent discrepancy.  If the abundance of Ni is important for an investigation, it is recommended to use the latest versions of {\texttt{AtomDB}} and {\texttt{SPEX}} and to note the current level of agreement between the models.

\begin{figure*}
\begin{center}
 \includegraphics[clip, trim =0mm 0mm 0mm 0mm, width=8.5cm,angle=0]{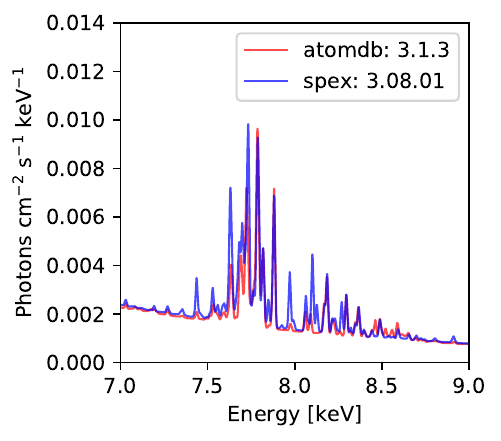} 
\includegraphics[clip, trim = 0mm 0mm 0mm 0mm, width=8.5cm,angle=0]{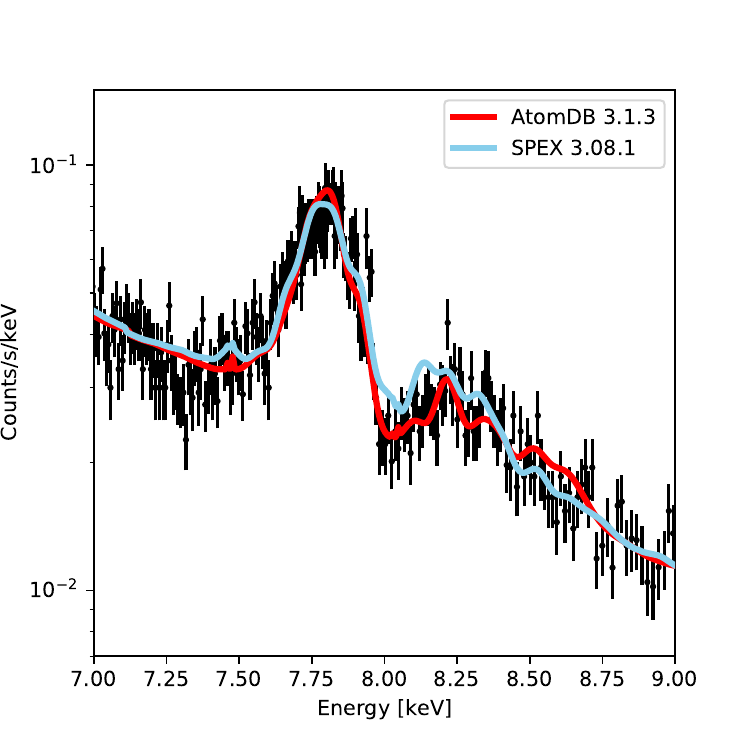} 
\end{center}
\caption{LEFT: Model spectra from AtomDB~3.1.3 and {\texttt{SPEX}}~3.08.01$^*$ in the 7.0.-9.0~keV range with parameters appropriate for the SE super-pixel ``e'' spectrum from Table~\ref{tab:spec_se} RIGHT: Data and model from SE super-pixel ``e'' fit with AtomDB~3.1.3 and {\texttt{SPEX}}~3.08.01$^*$ in the 7.0.-9.0~keV range. 
{Alt text: Two line graphs.  The left graph shows two lines with the model spectra from AtomDB 3.1.0 and {\texttt{SPEX}} 3.08.01$^*$ in the 7.0-9.0 keV band.  The horizontal axis shows the energy in keV. The vertical axis shows the flux in ${\mathrm {photons~cm^{-2}~s^{-1}~keV^{-1}}}$.  The right graph shows two lines with the models folded through the Resolve response. The horizontal axis  hows the energy in keV. The vertical axis shows the flux in ${\mathrm {counts~~s^{-1}~keV^{-1}}}$.}
}
\label{fig:fe_ni_atomdb_spex}
\end{figure*}

\subsection{The 3.2~keV feature}
\label{sec:3.2_feature}

\begin{figure}[!h]
\begin{center}
 \includegraphics[clip, trim =0mm 0mm 0mm 0mm, width=6.5cm,angle=0]{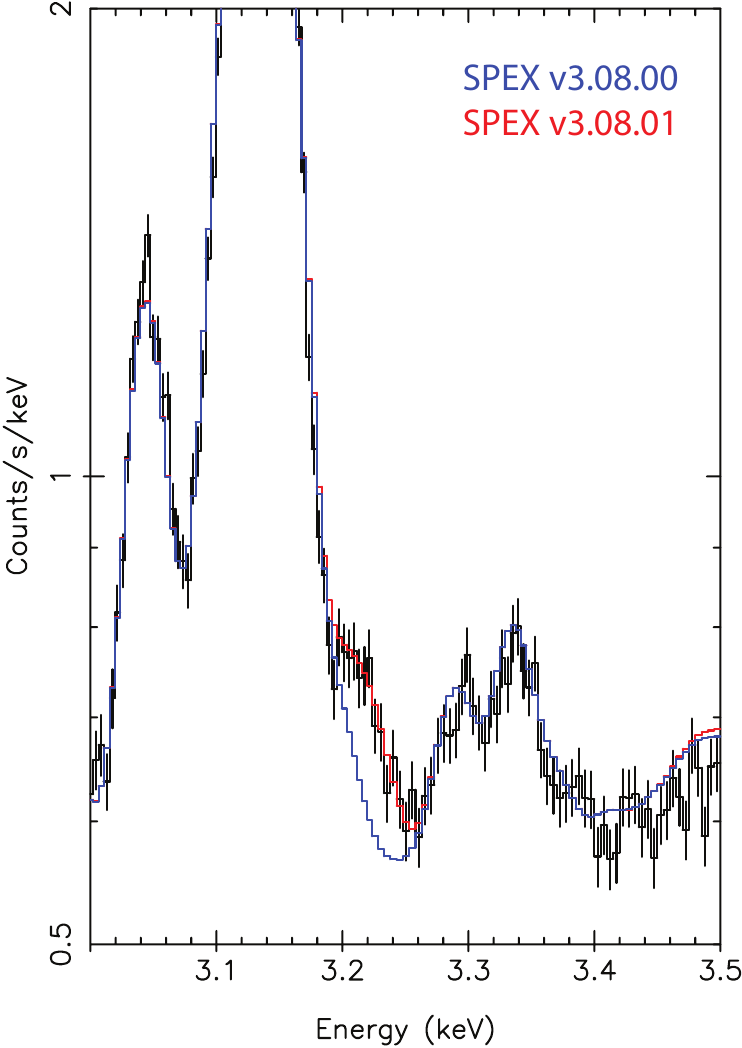} 
\end{center}
\caption{Comparison of {\texttt{SPEX}} versions 3.08.00 and 3.08.01$^*$ around 3.2 keV reveals a clear difference: {\texttt{SPEX}} 3.08.01$^*$ calculates the S XV Rydberg series up to $n$ = 52, significantly extending beyond the $n$ = 16 limit in {\texttt{SPEX}} 3.08.00. {\texttt{SPEX}} 3.08.01$^*$ provides a better match to the observational data in this energy range.
{Alt text: A line graph with two lines plotted on the data. The two lines show the spectral model with {\texttt{SPEX}}~3.08.00 and ~3.08.01$^*$. The horizontal axis shows the energy in keV. The vertical axis shows the flux in ${\mathrm {counts~~s^{-1}~keV^{-1}}}$.}

}
\label{fig:3.2keV}
\end{figure}

In the initial fits to the Cas~A spectra with {\texttt{AtomDB}}~3.0.9 and SPEX~3.08.00 there was a clear excess of emission around an energy of $\sim3.2$~keV that could not be explained by the models.
If this excess were real, it could potentially be radiative recombination continuum or a sign of charge exchange. However, as shown in Figure~\ref{fig:3.2keV}, the excess disappears when using the updated atomic database in {\texttt{SPEX}}~3.08.01$^*$.  The excess also disappears when {\texttt{AtomDB}}~3.1.0 is used. This suggests that the excess observed with {\texttt{AtomDB}}~3.0.9 and SPEX~3.08.00 is likely due to a collection of Rydberg series of S~XV with $n > 16$ that have been newly included in {\texttt{AtomDB}}~3.1.0 and SPEX~3.08.01$^*$.
These updated databases are important to support searches for relatively weak line emission from the odd-Z and trace elements such as P, Cl, \& K.
\\

\subsection{Element Abundances}
\label{sec:mn_abundance}

The \xrism\, Resolve spectra represent the highest-resolution, most sensitive spectra acquired to date in the 3.0-9.0~keV bandpass which contains emission from Ar, K, Ca, Sc, Ti, V, Cr, Mn, Fe, Co, \& Ni.  Therefore, the Resolve spectra provide useful constraints on the abundances of these elements that can be compared to nucleosynthesis models.   The amounts of these elements that are produced in the star and the subsequent explosion depend on many factors, among them the mass and metallicity
of the progenitor, the mass cut between the ejecta envelope and the proto-neutron star, the delay time between core collapse and explosion, and the explosion mechanism itself \citep{thielemann1996,sukhbold2016}.
There is compelling evidence that the SN explosion that produced the remnant we observe was asymmetric, see \cite{orlando2021} and references therein and \cite{suzuki2025,bamba2025,vink2025} in this volume.  The production of the odd-Z elements may be enhanced in explosive nucleosynthesis depending on the details of the explosion. Given the asymmetric nature of the Cas~A explosion, spatial variations in the ratios of the even-Z and and odd-Z elements might provide insights into the different nucleosynthesis processes and the stages at which they occur.

 The two regions in Cas~A examined in this paper provide a case study of the abundance ratios that may be used to inform nucleosynthesis models.  The ratio of the elements Ar, K, Ca, Ti, Cr, Mn,  \& Ni with respect to Fe are listed in Table~\ref{tab:abund_ratios} for the SE and NW spectra determined from the fits with {\texttt{AtomDB}}~3.1.0 and SPEX~3.08.01$^*$. All abundance ratios agree to within $1.0~\sigma$ between {\texttt{AtomDB}}~3.1.0 and SPEX~3.08.01$^*$ except for the Ni/Fe ratio for the SE spectrum which agrees to within $2.0~\sigma$. Ti and Mn are detected in the SE spectrum as previously reported by CCD observations \citep{sato2021,2023ApJ...954..112S} but their abundances are consistent with zero at the $1.0~\sigma$ level in the NW spectrum.  Ni is detected in both the SE and NW spectra, however, its abundance is consistent with zero at the $2.0~\sigma$ level in the NW spectrum. The abundance patterns are clearly different between the SE and NW spectra as the Ti/Fe, Mn/Fe, \& Ni/Fe ratios are significantly lower in the NW spectrum while the Ar/Fe, Ca/Fe, \& Cr/Fe ratios are comparable or perhaps slightly higher. These different abundance patterns may suggest that the innermost physical parameters (i.e., electron fraction and entropy) were inhomogeneous at the time of the explosion \citep{2018ApJ...852...40W,2023ApJ...954..114W}, consistent with the other evidence for an asymmetric explosion. A detailed analysis of these abundance ratios and the implications for nucleosynthesis models will be presented in future papers \citep{sato2025a,sato2025b}.

\begin{longtable}{lcccc}
  \caption{Element Abundance Ratios with respect to Fe with AtomdDB and SPEX
  }\label{tab:abund_ratios}  
\hline\noalign{\vskip3pt} 
  Abundance  & SE spectrum    &                    & NW spectrum &  \\   [2pt] 
  Ratio      & AtomDB 3.1.0 & SPEX 3.08.01$^*$ & AtomDB 3.1.0 & SPEX 3.08.01$^*$ \\   [2pt] 
\hline\noalign{\vskip3pt} 
\endfirsthead      
\hline\noalign{\vskip3pt}
 Abundance Ratio & SE spectrum    &                    & NW spectrum &  \\   [2pt] 
\hline\noalign{\vskip3pt} 
\endhead
\hline\noalign{\vskip3pt} 
\endfoot
\hline\noalign{\vskip3pt} 
\endlastfoot 
Ar/Fe & $0.65^{+0.06}_{-0.10}$   & $0.74^{+0.09}_{-0.04}$   & $0.84^{+0.11}_{-0.12}$   &  $1.00^{+0.18}_{-0.15}$ \\
K/Fe  & $0.86^{+0.13}_{-0.17}$   & $1.04^{+0.15}_{-0.15}$   & $0.97^{+0.33}_{-0.33}$   &  $1.13^{+0.41}_{-0.40}$ \\
Ca/Fe & $0.72^{+0.07}_{-0.11}$   & $0.84^{+0.02}_{-0.05}$   & $1.03^{+0.13}_{-0.15}$   &  $1.21^{+0.14}_{-0.16}$ \\
Ti/Fe & $0.90^{+0.28}_{-0.29}$   & $0.88^{+0.33}_{-0.30}$   & $0.23^{+0.74}_{-0.23}$   &  $0.17^{+0.66}_{-0.17}$ \\
Cr/Fe & $0.50^{+0.08}_{-0.10}$   & $0.63^{+0.10}_{-0.10}$   & $0.83^{+0.23}_{-0.23}$   &  $0.91^{+0.26}_{-0.27}$ \\
Mn/Fe & $0.50^{+0.13}_{-0.14}$   & $0.73^{+0.19}_{-0.19}$   & $0.14^{+0.36}_{-0.14}$   &  $0.35^{+0.56}_{-0.35}$ \\
Ni/Fe & $1.06^{+0.14}_{-0.19}$   & $0.64^{+0.09}_{-0.09}$   & $0.27^{+0.20}_{-0.19}$   &  $0.14^{+0.16}_{-0.14}$ \\
  \hline 
  \end{longtable}

\subsection{Semi-empirical Models for Calibration}
\label{sec:cal}

Cas~A has been used extensively by the current generation of X-ray instruments that use CCDs on \chandra, \xmm, Suzaku, and Swift as a calibration source given its line-rich spectrum and high flux. These CCD instruments are not as sensitive to small energy shifts (a few~eV) as Resolve and are almost entirely insensitive to broadening and asymmetry in the line shapes.  \cite{willingale2002} (hereafter W02) used the MOS instrument on \xmm\, to measure the distribution of redshifts throughout the remnant and  \cite{lazendic2006} (hereafter L06) used the high resolution spectra from the {\em{High Energy Tranmission Grating}} (HETG) on \chandra\, to measure the redshifts of several bright knots.  The companion papers in this volume \citep{suzuki2025,bamba2025,vink2025} expand upon this analysis by measuring not only the redshifts/blueshifts but also the widths and line shapes with the precision afforded by Resolve.  
W02 report a statistical error of $\sim3\%$ on the centroid of the Fe~{\footnotesize{XXV}}~He$\alpha$ triplet based on XMM MOS data, compared to the $0.3\%$ that Resolve achieved during these Cas~A observations. W02  note a dispersion of 24.2 eV on the Fe~{\footnotesize{XXV}}~He$\alpha$ centroid for the regions they analyzed which corresponds to a velocity of $1115~{\mathrm {km~s^{-1}}}$.  The primary cause of this dispersion is the fact that the centroid of the Fe~{\footnotesize{XXV}}~He$\alpha$  triplet varies as a function of temperature and ionization timescale, which the CCD instruments are not able to constrain well. In addition, W02  assumed zero width for the lines in their spectral fits. Bamba et al. 2025~(hereafter B25) report absolute values of the redshifts for the Fe~{\footnotesize{XXV}}~He$\alpha$ triplet from 1400 to $2160~{\mathrm {km~s^{-1}}}$ with an uncertainty of a few hundred ${\mathrm {km~s^{-1}}}$ based on the Resolve data.  B25 find that the broadening varies from 500 to $3000~{\mathrm {km~s^{-1}}}$ with the largest values occurring in the center of the remnant. L06 report absolute values of the redshift of the Si~{\footnotesize{XIII}}~He$\alpha$ triplet of 17 bright knots in Cas~A ranging from 380 to $4100~{\mathrm {km~s^{-1}}}$ with uncertainties of $2.5\%$ to $24\%$, corresponding to velocities of 70 to $650~{\mathrm {km~s^{-1}}}$ based on Chandra HETG data. L06 assumed a fixed width for the lines in their spectral fits. Suzuki et al. 2025~(hereafter S25) report absolute values of the redshift of the Si~{\footnotesize{XIII}}~He$\alpha$ triplet from the Resolve SE and NW pointings ranging from 480 to 1200 km/s  with uncertainties of $4.0\%$ to $10\%$, corresponding to velocities of 50 to $130~{\mathrm {km~s^{-1}}}$. S25 find that the broadening varies from 1240 to $1930~{\mathrm {km~s^{-1}}}$  with a typical uncertainty of $4\%$ or $\sim60~{\mathrm {km~s^{-1}}}$. The inclusion of broadening is essential in order to achieve acceptable fits with the Resolve data, while the CCD data are insensitive to the broadening.
Improving the fidelity of spectral models for sources like Cas~A has been a key objective of the {\em{International Astronomical Consortium for High Energy Calibration}}  ({\em{IACHEC}}, https://iachec.org), such as the standard model for the SNR 1E~0102.2-7219 \citep{plucinsky2017}.  The Resolve data promise to significantly improve the fidelity of the calibration models by determining the redshifts/blueshifts (energy shifts) to high precision and broadenings as a function of energy and position within Cas~A.  

The models using {\texttt{AtomDB}}~3.1.0 and {\texttt{SPEX}}~3.08.01$^*$ presented in Tables~\ref{tab:spec_se} \&~\ref{tab:spec_nw} include energy shifts as redshifts \& blueshifts and broadening of the lines in the two thermal components.  In addition, the {\texttt{bvvpshock}} model can produce asymmetric line profiles depending on the parameters. This effect is noticeable for the Fe-K complex as noted earlier. This information is crucial if one uses the spectra from Cas~A to calibrate the gain and/or spectral redistribution function of CCD instruments.  An important result described in \cite{suzuki2025} is that the redshift/blueshift for the 
He$\alpha$ complexes is different than for the Ly$\alpha$ lines of Si and S. Typically when fitting CCD-resolution spectra, the same redshift/blueshift was assumed for the He$\alpha$ and Ly$\alpha$ lines of the same element. Since the He$\alpha$ triplet usually has more counts than the Ly$\alpha$ lines, the fit would be driven by providing the best fit to the the He$\alpha$ triplet while sacrificing the fit for the  Ly$\alpha$ line.  This is further complicated by the fact that the CCD instruments cannot separate the forbidden, intercombination, and resonance lines of the triplets and the centroid of the triplet depends on the temperature and ionization timescale. This introduces a model dependence into the determination of the centroid of the triplets which is degenerate with the redshift/blueshift. Similarly, the spatial variation of the redshifts and blueshifts as described in \cite{bamba2025} and \cite{vink2025} needs to be included in a calibration model if spectra from different regions are to be analyzed.  In addition, \cite{vink2025} demonstrate that the shapes of the Ly$\alpha$ and He$\beta$ lines of Si and S are clearly non-Gaussian and vary with position in Cas~A. Typically when fitting CCD-resolution spectra, these lines were assumed to have a simple Gaussian or Lorentzian shapes, introducing another source of systematic error. For all of these reasons, an accurate determination of the line centroids and widths from a high spectral resolution instrument such as Resolve is beneficial for the development of standard models for calibration. 

The models presented in this paper should be an improvement over the existing models used to calibrate CCD instruments 
however, there are limitations that affect how they should be used. These models are ``semi-empirical'' in that they use the astrophysical plasma emission models in {\texttt{AtomDB}} and {\texttt{SPEX}}, a powerlaw model, and a detector background model as part of the NXB model.  The two plasma models, {\texttt{bvvrnei}} and {\texttt{bvvpshock}}, were selected because they adequately represent the data in the broad band, but caution should be applied in the physical interpretation of the parameters.  The plasmas along a given line of sight through Cas~A undoubtedly have a distribution of conditions (temperature, ionization timescale, abundances, etc.) that we are representing with the combination of two models, each with its own limited set of parameters. In addition,  these plasma emission models will change if the underlying data in the {\texttt{AtomDB}} and {\texttt{SPEX}} databases change.  A standard {\em{IACHEC}} model typically freezes the version of {\texttt{AtomDB}} and {\texttt{SPEX}} to be used with it to eliminate this issue. Finally, the relatively large PSF of the \xrism\, mirrors will mix emission from multiple regions as described in \ref{sec:ssm}, complicating the comparison to instruments with smaller PSFs such as \chandra\, and \xmm.  
The broad band spectra are well-fit by these models, but there are issues around some of the line complexes that produce significant residuals in the fits.  Figure~\ref{fig:se_s_ar_fe} presents a closeup view of the data and model for the S~{\footnotesize{XV}}, Ar~{\footnotesize{XVII}}, and Fe~{\footnotesize{XXV}} He$\alpha$ triplets.  Note that the two thermal components contribute roughly equally to the S~{\footnotesize{XV}}~He$\alpha$ triplet, the high temperature component provides the majority of the Ar~{\footnotesize{XVII}}~He$\alpha$ triplet, and the high temperature component provides essentially all of the Fe~{\footnotesize{XXV}}~He$\alpha$ triplet.  The combination of the two models, each with its own redshift/blueshift and broadening, has several degrees of freedom to produce a complex line shape to match the data. 
The two component thermal model accurately reproduces the centroids of the line complexes but is not able to capture the complex shape of the S~{\footnotesize{XV}} and Fe~{\footnotesize{XXV}} He$\alpha$ triplets, but it does reasonably well for Ar~{\footnotesize{XVII}}. Therefore, these models should be sufficient for the calibration of the gain and spectral redistribution function of CCD-resolution instruments, as long as the highest fidelity is not needed for the spectral redistribution function. An alternative approach is to fit in a narrower energy range and focus on just a few lines of interest as was done in \cite{suzuki2025,bamba2025,vink2025}.

\begin{figure*}
\begin{center}
\includegraphics[clip, trim =5mm 0mm 40mm 25mm, width=5.8cm,angle=0]
{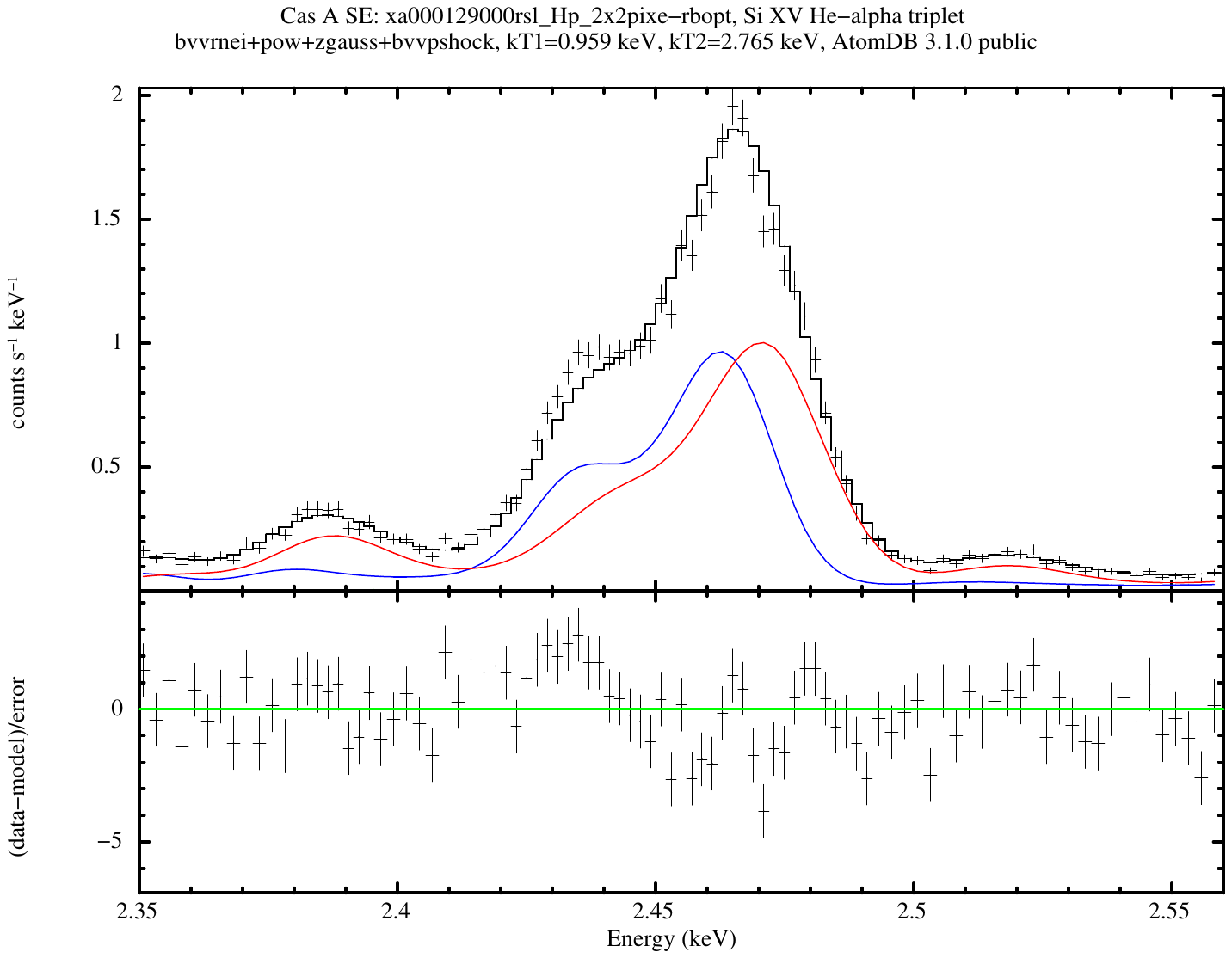} 
\includegraphics[clip, trim =5mm 0mm 40mm 25mm, width=5.8cm,angle=0]
{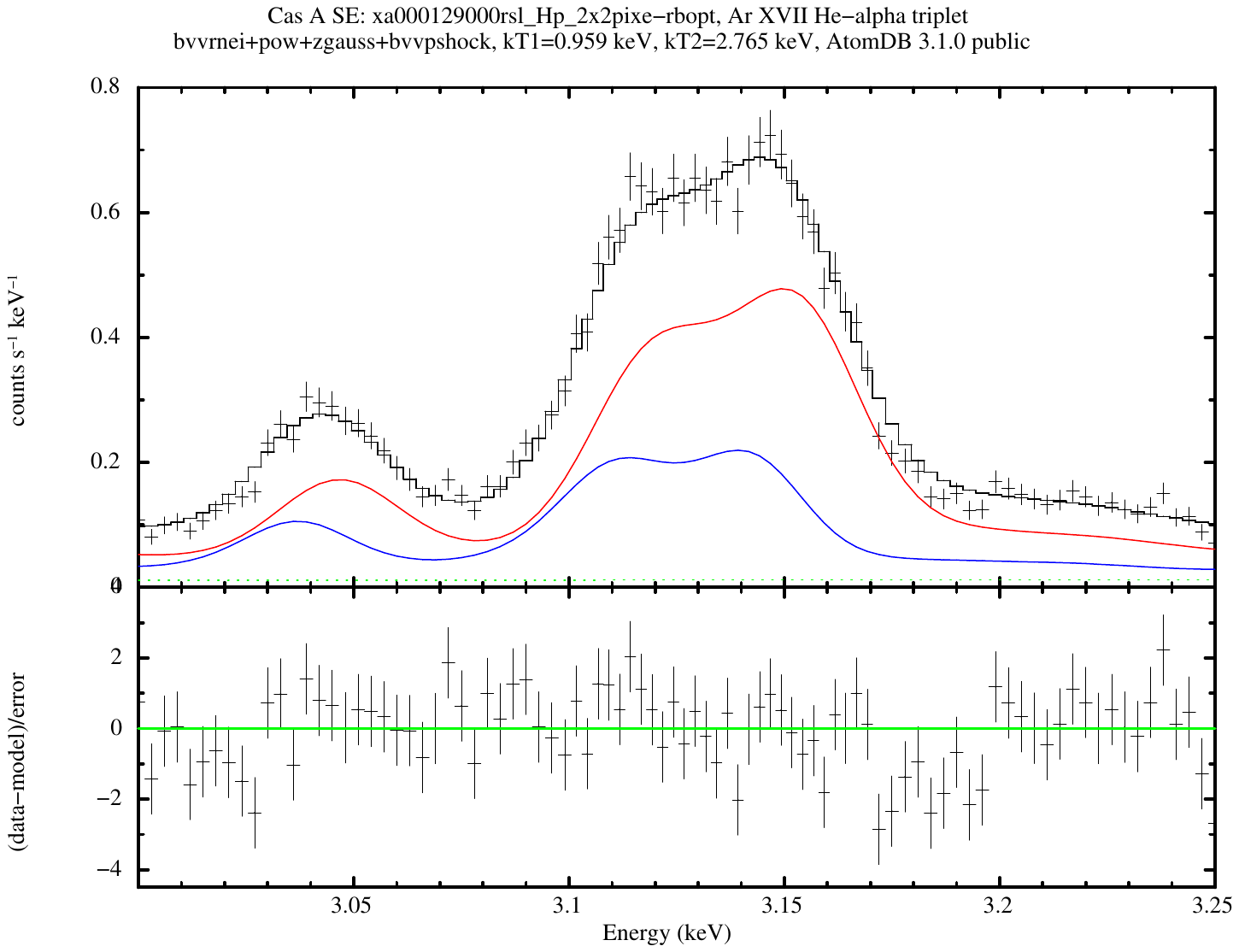} 
\includegraphics[clip, trim =5mm 0mm 40mm 25mm, width=5.8cm,angle=0]
{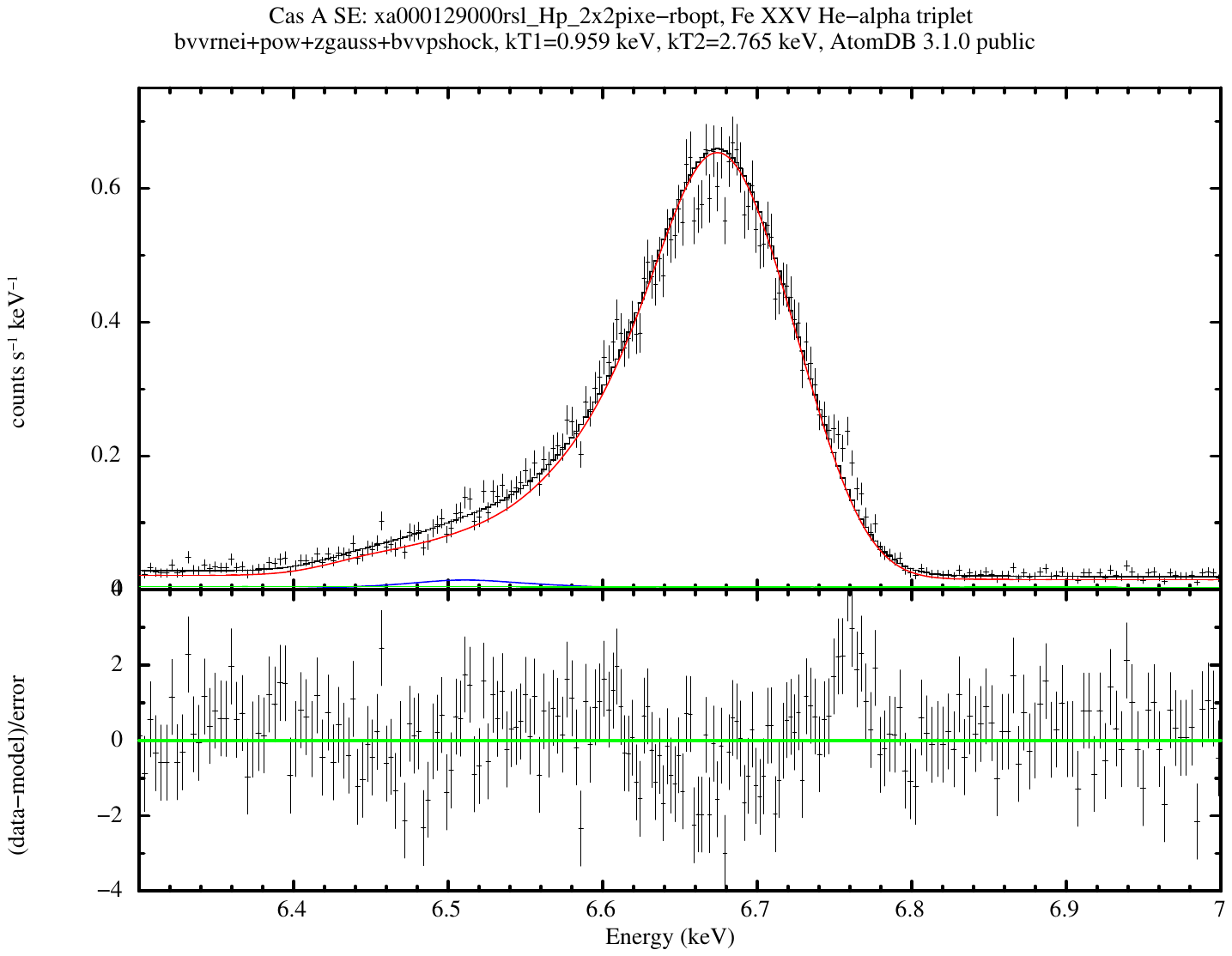} 

\end{center}
\caption{Resolve spectra from the on-axis point, so-called super-pixel ``e'', for the SE pointing fit with XSPEC and AtomDB~3.1.0 in the S~{\footnotesize{XV}} He$\alpha$ energy range (LEFT), the Ar~{\footnotesize{XVII}} He$\alpha$ energy range (CENTER), and the Fe~{\footnotesize{XXV}} He$\alpha$ energy range (RIGHT).  The data and total model are plotted in black, the low temperature {\texttt{bvvrnei}} component in blue and the high temperature {\texttt{bvvpshock}} component in red.
{Alt text: Three line graphs.  Each graph has three lines plotted on the data. The three lines show the total spectral model, the low temperature {\texttt{bvvrnei}} component, and the high temperature  {\texttt{bvvpshock}} component with AtomDB~3.1.0. The horizontal axis  shows the energy in keV. The vertical axis shows the flux in ${\mathrm {counts~~s^{-1}~keV^{-1}}}$.}
}
\label{fig:se_s_ar_fe}
\end{figure*}

\section{Conclusion}
\label{sec:conclusion}

We have presented an overview of two observations of the Cas~A SNR conducted during the PV phase of the \xrism\, mission. One observation was centered at the bright region in the SE and the other in the NW. We presented images in the 1.5-10.0 keV band from the Xtend and Resolve instruments that identify the regions of the remnant covered by the Resolve instrument.  We quantified the \xrism\, SSM effect by generating maps in detector pixels of the percentage of events that originate from the corresponding sky pixel region (so-called ``purity'' maps) and by generating two-dimensional matrices of the contribution of each pixel to every other pixel in the array.  These calculations include the PSF effects and the attitude drifts during the observations. We find that the highest purity value for a single super-pixel ($2\times2$ pixels) is $48\%$ while the lowest is $9\%$, highlighting the importance of the SSM effect for an extended object with a complex morphology like Cas~A.  We extracted spectra from a $2\times2$ pixel region on-axis from the SE and NW observations to minimize the effects of SSM and uncertainties in the off-axis response of the mirror. We developed a semi-empirical model consisting of two thermal components ({\texttt{bvvrnei}} and {\texttt{bvvpshock}}), a nonthermal component (a powerlaw) and detector background that fits the spectra well from 1.5-15.0~keV.  We found that it is essential to incorporate redshifts/blueshifts and broadening of the emission lines to achieve a reasonable fit given the high spectral resolution of Resolve.

We fit these spectra with the versions of {\texttt{AtomDB}} and {\texttt{SPEX}} that were available just before the launch of \xrism\, and the versions that are available as of the writing of this paper. We described the changes made in the post-launch versions and documented the differences in the fitted parameters comparing the pre- and post-launch versions.  The most significant difference is the abundance of Ni in the {\texttt{AtomDB}} models, which can be lower by as much as a factor of three in the post-launch version depending on the spectral model.   The agreement of the fitted parameters between the post-launch versions of {\texttt{AtomDB}} and {\texttt{SPEX}} is excellent with most parameters agreeing to within $1.0\sigma$, except for the Ni abundance, the redshifts, and broadening. The most significant disagreement between {\texttt{AtomDB}} and {\texttt{SPEX}} is the abundance of Ni. The redshift and broadening parameters are assuming different values in the fits with {\texttt{AtomDB}} and {\texttt{SPEX}} such that the combination reproduces the widths of the lines. Therefore, the differences in the individual parameters (redshift and broadening) do not manifest themselves as a large difference in the model spectra. We found that the abundance ratios of Ti/Fe, Mn/Fe, \& Ni/Fe are significantly lower in the NW region than in the SE region and the abundances of Ti and Mn are consistent with zero in the NW region.  This large variation in the abundances of these elements is further evidence of an asymmetric explosion that affected the nucleosynthesis processes.  Finally, we suggested that the models presented in this paper would be useful for the calibration of CCD-resolution instruments but described the limitations of their use.

\begin{ack}
We thank all of the scientists, engineers, and technicians that built the \xrism\, satellite, operate the mission, and developed the software that we used to analyze the data in this paper. We thank Hiromasa Suzuki, who acted as the internal referee, and the anonymous referee for helpful comments.
PP acknowledges support from NASA \xrism\, grants 80NSSC18K0988 and 80NSSC23K1656, and the Smithsonian Institution and the Chandra X-ray Center through NASA contract NAS8-03060.
MA and JV acknowledge financial support from NWO under grant number 184.034.002.
This work was partly supported by Japan Society for the Promotion of
Science Grants-in-Aid for Scientific Research (KAKENHI) Grant Numbers,
JP23K25907 (AB).
This work was supported by the JSPS Core-to-Core Program (grant number: JPJSCCA20220002) (YT,HS) and the Japan Society for the Promotion of Science Grants-in-Aid for Scientific Research (KAKENHI) Grant Number JP20K04009 (YT).

\end{ack}


\begin{thebibliography}{}
%
%
\bibitem[Arnaud(1996)]{arnaud1996}
Arnaud, K.~A.\ 1996, Astronomical Data Analysis Software and Systems V, 101, 17
%
\bibitem[Bamba et al.(2025)]{bamba2025}
Bamba, A., et al.\ 2025, \pasj, arXiv e-prints, arXiv:2504.03268
%
\bibitem[Bryans et al.(2009)]{Bryans2009}
Bryans, P., Landi, E., Savin, D.~W.\ 2009, \apj, 691, 1540
%
\bibitem[Cash(1979)]{cash1979}
Cash, W.\ 1979, \apj, 228, 939
%
\bibitem[Chevalier \& Oishi(2003)]{chevalier2003} 
Chevalier, R.~A. \& Oishi, J.\ 2003, \apjl, 593, L23
%
\bibitem[DeLaney et al.(2010)]{delaney2010}
DeLaney, T., Rudnick, L., Stage, M.~D., et al.\ 2010, \apj, 725, 2038
%
\bibitem[DeLaney et al.(2014)]{delaney2014} DeLaney, T., Kassim, N.~E., Rudnick, L., et al.\ 2014, \apj, 785, 7
%
\bibitem[Fesen et al.(2006)]{fesen2006}
Fesen, R.~A., Hammell, M.~C., Morse, J., et al.\ 2006, \apj, 645, 283 
%
\bibitem[Fesen \& Milisavljevic(2016)]{fesen2016} Fesen, R.~A. \& Milisavljevic, D.\ 2016, \apj, 818, 1, 17 
%
\bibitem[Foster et al.(2012)]{foster2012}
Foster, A.~R., Ji, L., Smith, R.~K., et al.\ 2012, \apj, 756, 128
%
\bibitem[Fr{\"o}hlich et al.(2006)]{2006ApJ...637..415F} Fr{\"o}hlich, C., Hauser, P., Liebend{\"o}rfer, M., et al.\ 2006, \apj, 637, 1, 415
%
\bibitem[Fujimoto \& Nagakura(2019)]{2019MNRAS.488L.114F} Fujimoto, S.-. ichiro . \& Nagakura, H.\ 2019, \mnras, 488, 1, L114 
%
\bibitem[Grefenstette et al.(2014)]{grefenstette2014} 
Grefenstette, B.~W., Harrison, F.~A., Boggs, S.~E., et al.\ 2014, \nat, 506, 339
%
\bibitem[Hahn et al.(2017)]{Hahn2017Electron-impactZinc}
    Hahn, M., M{\"{u}}ller, A., Savin, D.~W.\ 2017, \apj, 850, 122
%
\bibitem[Hayashi et al.(2024)]{hayashi2024} Hayashi, T., Boissay-Malaquin, R., Tamura, K., et al.\ 2024, \procspie, 13093, 130931L 
%
\bibitem[Helder \& Vink(2008)]{helder2008}
Helder, E.~A. \& Vink, J.\ 2008, \apj, 686, 1094
%
\bibitem[Holt et al.(1994)]{holt1994} Holt, S.~S., Gotthelf, E.~V., Tsunemi, H., et al.\ 1994, \pasj, 46, L151. 
%
\bibitem[Hughes et al.(2000)]{hughes2000}
Hughes, J.~P., Rakowski, C.~E., Burrows, D.~N., et al.\ 2000, \apjl, 528, L109
%
\bibitem[Hwang et al.(2004)]{hwang2004} Hwang, U., Laming, J.~M., Badenes, C., et al.\ 2004, \apjl, 615, 2, L117
%
\bibitem[Hwang \& Laming(2012)]{hwang2012} Hwang, U. \& Laming, J.~M.\ 2012, \apj, 746, 130
%
\bibitem[Ishisaki et al.(2022)]{ishisaki2022}
Ishisaki, Y., Kelley, R., Awaki, H., et al.\ 2022,\procspie,  12181, 121811S
%
\bibitem[Jerkstrand et al.(2015)]{jerkstrand2015} Jerkstrand, A., Timmes, F.~X., Magkotsios, G., et al.\ 2015, \apj, 807, 1, 110. 
%
\bibitem[Kaastra et al.(1996)]{kaastra1996} Kaastra, J.~S., Mewe, R., \& Nieuwenhuijzen, H.\ 1996, UV and X-ray Spectroscopy of Astrophysical and Laboratory Plasmas, 411
%
\bibitem[Kaastra \& Bleeker(2016)]{kaastra2016}
Kaastra, J.~S. \& Bleeker, J.~A.~M.\ 2016, \aap, 587, A151 
%
\bibitem[Krause et al.(2008)]{krause2008} 
Krause, O., Birkmann, S.~M., Usuda, T., et al.\ 2008, Science, 320, 1195
%
\bibitem[Laming \& Hwang(2003)]{laming2003} 
Laming, J.~M. \& Hwang, U.\ 2003, \apj, 597, 347
%
\bibitem[Laming \& Temim(2020)]{laming2020} 
Laming, J.~M. \& Temim, T.\ 2020, \apj, 904, 115 
%
\bibitem[Lazendic et al.(2006)]{lazendic2006}
Lazendic, J.~S., Dewey, D., Schulz, N.~S., et al.\ 2006, \apj, 651, 250 
%
\bibitem[Lee et al.(2014)]{lee2014} 
Lee, J.-J., Park, S., Hughes, J.~P., et al.\ 2014, \apj, 789, 7
%
\bibitem[Lodders et al.(2009)]{lodders2009} Lodders, K., Palme, H., \& Gail, H.-P.\ 2009, Landolt B{\"o}rnstein, 4B, 712
%
\bibitem[Markert et al.(1983)]{markert1983} Markert, T.~H., Canizares, C.~R., Clark, G.~W., et al.\ 1983, \apj, 268, 134 
%
\bibitem[Milisavljevic \& Fesen(2013)]{milisavljevic2013}
Milisavljevic, D. \& Fesen, R.~A.\ 2013, \apj, 772, 134
%
\bibitem[Mori et al.(2022)]{mori2022} Mori, K., Tomida, H., Nakajima, H., et al.\ 2022, \procspie, 12181, 121811T
%
\bibitem[Noda et al.(2025)]{noda2025}
Noda, H., et al.\ 2025, \pasj, arXiv e-prints, arXiv:2503.06760
%
%
\bibitem[Orlando et al.(2016)]{orlando2016} 
Orlando, S., Miceli, M., Pumo, M.~L., et al.\ 2016, \apj, 822, 22 
%
\bibitem[Orlando et al.(2021)]{orlando2021}
Orlando, S., Wongwathanarat, A., Janka, H.-T., et al.\ 2021, \aap, 645, A66
%
\bibitem[Orlando et al.(2022)]{orlando2022}
Orlando, S., Wongwathanarat, A., Janka, H.-T., et al.\ 2022, \aap, 666, A2
%
\bibitem[Patnaude et al.(2011)]{patnaude2011} Patnaude, D.~J., Vink, J., Laming, J.~M., et al.\ 2011, \apjl, 729, L28
%
\bibitem[Plucinsky et al.(2017)]{plucinsky2017} Plucinsky, P.~P., Beardmore, A.~P., Foster, A., et al.\ 2017, \aap, 597, A35
%
\bibitem[Porter et al.(2024)]{porter2024}
Porter, F.~S., Kilbourne, C.~A., Chiao, M., et al.\ 2024, 
\procspie, 13093, 130931K
%
\bibitem[Reed et al.(1995)]{reed1995}
Reed, J,~E., Hester, J.~J., Fabian, A. C., et al.\ 1995, \apj, 440, 708
%
\bibitem[Rest et al.(2011)]{rest2011} 
Rest, A., Foley, R.~J., Sinnott, B., et al.\ 2011, \apj, 732, 3. doi:10.1088/0004-637X/732/1/3
%
\bibitem[Sato et al.(2021)]{sato2021}
Sato, T., Maeda, K., Nagataki, S., et al.\ 2021, \nat, 592, 537
%
\bibitem[Sato et al.(2023)]{2023ApJ...954..112S} Sato, T., Yoshida, T., Umeda, H., et al.\ 2023, \apj, 954, 2, 112

%
\bibitem[Sato et al.(2025a)]{sato2025a}
Sato, T. et al.\ 2025, \nat, submitted
%
%
\bibitem[Sato et al.(2025b)]{sato2025b}
Sato, T. et al.\ 2025, \apj, in preparation
%
\bibitem[Sukhbold et al.(2016)]{sukhbold2016} Sukhbold, T., Ertl, T., Woosley, S.~E., et al.\ 2016, \apj, 821, 1, 38
%
\bibitem[Suzuki et al.(2025)]{suzuki2025}
Suzuki, S., et al.\ 2025, \pasj, arXiv e-prints, arXiv:2503.23640
%
%
\bibitem[Tashiro et al.(2020)]{tashiro2020}
Tashiro, M., Maejima, H., Toda, K., et al.\ 2020, \procspie, 11444, 1144422
%
\bibitem[Tashiro et al.(2024)]{tashiro2024} Tashiro, M., Watanabe, S., Maejima, H., et al.\ 2024, \procspie, 13093, 130931G 
%
\bibitem[Tashiro et al.(2025)]{tashiro2025} Tashiro, M., Kelley, R., Watanabe, S., et al.\ 2025, \pasj, doi:10.1093/pasj/psaf023
%
\bibitem[Thielemann et al.(1996)]{thielemann1996} Thielemann, F.-K., Nomoto, K., \& Hashimoto, M.-A.\ 1996, \apj, 460, 408 
%
\bibitem[Thorstensen et al.(2001)]{thorstensen2001}
Thorstensen, J.~R., Fesen, R.~A., \& van den Bergh, S.\ 2001, \aj, 122, 297 
%
\bibitem[Tsuchioka et al.(2022)]{tsuchioka2022} Tsuchioka, T., Sato, T., Yamada, S., et al.\ 2022, \apj, 932, 93
%
\bibitem[Uchida et al.(2025)]{uchida2025}
Uchida, H., et al.\ 2025, \pasj, arXiv e-prints, arXiv:2503.20180
%
%
\bibitem[Uchiyama \& Aharonian(2008)]{uchiyama2008} Uchiyama, Y. \& Aharonian, F.~A.\ 2008, \apjl, 677, L105 
%
\bibitem[Urdampilleta et al.(2017)]{Urdampilleta2017a}
    Urdampilleta, I., Kaastra, J.~S., Mehdipour, M.\ 2017, \aap, 601, A85
%
%
\bibitem[Vink et al.(2022)]{vink2022}
Vink, J., Patnaude, D.~J., \& Castro, D.\ 2022, \apj, 929, 57 
%
\bibitem[Vink et al.(2024)]{vink2024} Vink, J., Agarwal, M., Slane, P., et al.\ 2024, \apjl, 964, L11 
%
\bibitem[Vink et al.(2025)]{vink2025}
Vink, J., et al.\ 2025, \pasj, arXiv e-prints, arXiv:2505.04691
%

\bibitem[Wanajo et al.(2018)]{2018ApJ...852...40W} Wanajo, S., M{\"u}ller, B., Janka, H.-T., et al.\ 2018, \apj, 852, 1, 40 

\bibitem[Wang \& Burrows(2023)]{2023ApJ...954..114W} Wang, T. \& Burrows, A.\ 2023, \apj, 954, 2, 114


\bibitem[Willingale et al.(2002)]{willingale2002}
Willingale, R., Bleeker, J.~A.~M., van der Heyden, K.~J., Kaastra, J.~S., Vink, J.\ 2002, \aap, 381, 1039
%
\bibitem[Wongwathanarat et al.(2015)]{wongwathanarat2015}
Wongwathanarat, A., M{\"u}ller, E., \& Janka, H.-T.\ 2015, \aap, 577, A48
%
\bibitem[Wongwathanarat et al.(2017)]{wongwathanarat2017}
Wongwathanarat, A., Janka, H.-T., M{\"u}ller, E., et al.\ 2017, \apj, 842, 13
%
    

    
\end{thebibliography}

\end{document}